\documentclass{aa}
\usepackage[varg]{txfonts}
\usepackage{natbib,twoopt}
\bibpunct{(}{)}{;}{a}{}{,}             
\makeatletter
  \newcommandtwoopt{\citeads}[3][][]{\href{http://adsabs.harvard.edu/abs/#3}%
    {\def\hyper@linkstart##1##2{}%
     \let\hyper@linkend\@empty\citealp[#1][#2]{#3}}}
  \newcommandtwoopt{\citepads}[3][][]{\href{http://adsabs.harvard.edu/abs/#3}%
    {\def\hyper@linkstart##1##2{}%
     \let\hyper@linkend\@empty\citep[#1][#2]{#3}}}
  \newcommandtwoopt{\citetads}[3][][]{\href{http://adsabs.harvard.edu/abs/#3}%
    {\def\hyper@linkstart##1##2{}%
     \let\hyper@linkend\@empty\citet[#1][#2]{#3}}}
  \newcommandtwoopt{\citeyearads}[3][][]%
    {\href{http://adsabs.harvard.edu/abs/#3}
    {\def\hyper@linkstart##1##2{}%
     \let\hyper@linkend\@empty\citeyear[#1][#2]{#3}}}
\makeatother

\begin{document}

\title{CHANG-ES XII: A LOFAR and VLA view of the edge-on star-forming galaxy NGC 3556}
\author{A. Miskolczi \inst{1} 
\and V. Heesen \inst{2} 
\and C.~Horellou\inst{3}
\and D.-J. Bomans \inst{1}
\and R. Beck \inst{4}
\and G. Heald \inst{5}
\and R.-J. Dettmar \inst{1} 
\and S. Blex \inst{1} 
\and B. Nikiel-Wroczyński \inst{6}
\and K.T. Chy\.zy  \inst{6} 
\and Y. Stein \inst{7}
\and J. A. Irwin \inst{8}
\and T. W. Shimwell \inst{9}
\and Q. D. Wang \inst{10}
}

\institute{
Astronomisches Institut der Ruhr-Universit\"at Bochum, Universit\"atsstr. 150, D-44780 Bochum, Germany 
\and Universit\"at Hamburg, Hamburger Sternwarte, Gojenbergsweg 112, D-21029 Hamburg, Germany 
\and Chalmers University of Technology,
    Dept of Space, Earth and Environment, 
    Onsala Space Observatory, 
    SE-439 92 Onsala, Sweden 
\and Max-Planck Institut f\"ur Radioastronomie, Auf dem H\"ugel 69, D-53121 Bonn, Germany 
\and CSIRO Astronomy and Space Science, PO Box 1130, Bentley WA 6102, Australia  
\and Astronomical Observatory, Jagiellonian University, ul. Orla 171, Kraków PL 30-244, Poland 
\and Observatoire astronomique de Strasbourg, Université de Strasbourg, CNRS, UMR 7550, 11 rue de l’Université, 67000 Strasbourg, France 
\and Department of Physics, Engineering Physics, and Astronomy, Queen’s University, Kingston, Ontario K7L 3N6, Canada
\and ASTRON, the Netherlands Institute for Radio Astronomy, Postbus 2,7990 AA, Dwingeloo, The Netherlands 
\and Department of Astronomy, University of Massachusetts, 710 North Pleasant St, Amherst, MA 01003, USA
}

\abstract{Low-frequency radio continuum studies of star-forming edge-on galaxies can help to further understand how cosmic-ray electrons (CRe) propagate through the interstellar medium into the halo and how this is affected by energy losses and magnetic fields.}
{Observations with the Very Large Array (VLA) from Continuum Haloes in Nearby Galaxies -- an EVLA Survey (CHANG-ES) are combined with those with the LOw Frequency ARray (LOFAR) from the LOFAR Two-metre Sky Survey (LoTSS ) to identify the prevailing mode of cosmic-ray transport in the edge-on spiral galaxy NGC\,3556.}
{We mapped the radio spectral index, magnetic field strength, and orientation using VLA $1.5$  and 6 GHz and LOFAR 144 MHz data, and we fit 1D cosmic-ray propagation models to these maps using \textsc{Spinnaker} (Spectral Index Numerical Analysis of K(c)osmic-ray electron radio emission) and its interactive wrapper \textsc{Spinteractive}.}
{We find that the spectral index in the galactic midplane is, as expected for young CRe, $\alpha \approx -0.7$ and steepens towards the halo of the galaxy as a consequence of spectral ageing. 
The intensity scale heights are about $1.4$ and $1.9$~kpc for the thin disc, and $3.3$ and $5.9$~kpc for the thick disc at $1.5$~GHz and 144~MHz, respectively. While pure diffusion cannot explain our data, advection can, particularly if we assume a linearly accelerating wind. Our best-fitting model has an initial speed of 123~km~s$^{-1}$ in the galactic midplane and reaches the escape velocity at heights between 5~kpc and 15~kpc above the disc, depending on the assumed dark matter halo of the galaxy. This galactic wind scenario is corroborated by the existence of vertical filaments seen both in the radio continuum and in H\,$\alpha$ in the disc--halo interface and of a large-scale reservoir of hot, X-ray emitting gas in the halo.}
{Radio haloes show the existence of galactic winds, possibly driven by cosmic rays, in typical star-forming spiral galaxies.}

\maketitle 

\section{Introduction}

Studying edge-on galaxies opens up the possibility to observe something that is not feasible with face-on galaxies: the halo of the galaxy. It is now clear that sufficiently deep and well-resolved radio observations of edge-on galaxies show emission high above their galactic discs \citep{changes_dr1}. Often, the radio halo extent is greater than the radial extent, as observations of a large number of edge-on galaxies show. This can provide insight into the mechanisms that lead to the formation of radio haloes, in particular galactic winds, which are a key factor
in galaxy evolution \citep{Veilleux2005}. Stellar feedback is thought to play 
a significant role in galaxies below halo masses of $10^{12}~\rm M_\sun$, whereas active galactic nuclei (AGN) dominate the feedback process in more massive galaxies (e.g. \citealt{Silk2013, Li2018}). The details of stellar feedback are important, and the question of how much mass, heavier elements (metals), and angular momentum are removed over time is crucial to explain the observed characteristics of galaxies \citep{Scannapieco2002, Scannapieco2008}. In this context, cosmic-ray driven winds are important as they facilitate winds in normal (i.e. non-starbursting) late-type galaxies such as our own Milky Way. 
Cosmic-ray driven winds cannot only occur in conditions of low star formation rate (SFR) surface densities where purely thermally driven winds fail \citep{Everett2008}, but can also drive slower, cooler winds that are much denser than the hot, fast thermally driven winds \citep{Girichidis2018}. These winds shape in particular the lower mass galaxies, such as dwarf irregular galaxies \citep{Tremonti2004}, so it is important to understand how they work. 

Observationally, cosmic rays outside of our Milky Way can be studied via their electron component, the cosmic-ray electrons (CRe). These GeV electrons spiral around magnetic field lines and emit synchrotron radiation that has a characteristic spectral index of  $\alpha \approx -0.7$ ($I_\nu\propto \nu^{\alpha}$), which can be distinguished from a thermal radio continuum spectrum with a spectral index of $\alpha\approx -0.1$. 
Lower radio frequencies are beneficial to study radio haloes for two reasons. First, the influence of the thermal radio continuum can be neglected, which makes it possible to study the uncontaminated synchrotron emission component. Second, as a result of  spectral ageing, lower frequencies are more sensitive to older emission, which allows
us to study the emission farther away from the star formation regions in the disc, where young CRe are injected. These low-frequency studies have become possible with the advent of the LOw Frequency ARray \citep[LOFAR;][]{lofar2013}, which gives us a detailed view on galaxies at a frequency of about 140~MHz.

This work is the third study of a radio halo with LOFAR. 
\cite{Heesen2018} have observed the Local Group starburst dwarf irregular galaxy IC\,10 at 140~MHz and showed that the galaxy likely possesses a slow wind starting at only $20~\rm km\,s^{-1}$, possibly accelerating further into the halo and exceeding the escape velocity within 1~kpc from the disc. \cite{Mulcahy2018} have investigated the propagation of cosmic rays in the late-type spiral edge-on galaxy NGC\,891. Their 146 MHz map showed intensity scale heights that were 70\% larger than at $1.5$~GHz, which shows the promise of low-frequency observations in studying the halo structure of spiral galaxies. Without detailed modelling of cosmic-ray propagation, these authors could not decide whether diffusion or advection in a wind is responsible for the CRe in the halo. This is something we would like to address in this study of NGC\,3556.

NGC\,3556 (M\,108, UGC\,06225) is a star-forming, nearby edge-on spiral galaxy at a distance of 14~Mpc with an inclination of $81\degr$ \citep{irwin2012}. The galaxy has a  SFR of $2.17~\rm M_\odot\,yr^{-1}$ and a SFR surface density (SFRD) of $4.4\times 10^{-3}~\textrm{M}_\sun\,\textrm{yr}\,^{-1}\,\textrm{kpc}^{-2}$, which is typical for late-type edge-on galaxies \citep{changes_dr1}. 

Most studies of this galaxy have focussed on localised objects such as \ion{H}{I} shells \citep{King1997} or the disc--halo interaction \citep{wang2003}. An early radio continuum study showed that the spectral index of the galaxy steepens towards the halo \citep{debruyn1979}, which is often seen in other edge-on galaxies 
(i.e. NGC\,5775, \citealt{duric1998}; NGC\,4631, \citealt{Hummel1990}; NGC\,253, \citealt{heesen2009_253}). The favoured explanations of the steepening are either a cosmic-ray ageing effect, limiting the lifetime of the electrons to about $10^7$ years, or that the magnetic field is weaker in the halo than in the disc. The authors also note that a single-component fit cannot reproduce the vertical profile of the surface brightness of the radio continuum emission. 
The low-frequency capabilities of LOFAR make it possible to observe different components of the galaxy, such as low-energy CRe. These CRe are believed to originate in supernova remnants from diffusive shock acceleration \citep{Bell1978, Blandford1978, Caprioli2012}. This implies that a galaxy with a higher SFR and, consequently, a higher rate of exploding supernovae (SNe), produce more low-energy CRe that can be detected at low radio frequencies. 
Various energy-loss processes and transport mechanisms can change the expected vertical profile of a galaxy. A faster wind increases the scale height by driving the CRe out into, or even out of, the halo. 
This has been seen in extreme star-forming galaxies such as M\,82 \citep{adebahr2013}. On the other hand, strong energy losses may lead to the non-detection of CRe far out in the halo \citep{Beck2015}.

The possibility to detect CRe also depends on the presence of a galactic magnetic field, without which the electrons would not radiate synchrotron radiation. Studying the distribution of radio synchrotron emission and
the magnetic field helps us to constrain the parameters and the effects responsible for the observed properties of the halo.

In this paper, we present deep LOFAR observations of NGC\,3556. These data are combined with observations from the Continuum Haloes in Nearby Galaxies -- an EVLA Survey \citep[CHANG-ES;][]{irwin2012} to study the distribution of CRe and magnetic fields. This galaxy is particularly interesting to test whether late-type galaxies with SFRs similar to the Milky Way have outflows, in addition to the starburst galaxies mentioned earlier. This paper is organised as follows. In Sect.~\ref{sec:data}, we present the data used in the analysis. Section~\ref{sec:results} contains our results. 
The results are discussed in Sect.~\ref{sec:discussion}, and conclusions are presented in Sect.~\ref{sec:conclusions}.

\section{Data}
\label{sec:data}
\subsection{LoTSS}
\label{lotss}
The low-frequency images were taken from the LOFAR Two-metre Sky Survey \citep[LoTSS;][]{Shimwell2017}. This survey aims to cover the entire northern sky with a total of 3170 pointings, each being observed for eight hours. This produces fields with a resolution of 6 arcseconds and a noise level of about 100~$\mu$Jy/beam.
The LoTSS first data release covers a 424-square-degree region of the HETDEX Spring Field (right ascension 10h45m00s to 15h30m00s and declination $45\degr 00\arcmin 00\arcsec$ to $57\degr 00\arcmin 00\arcsec$), which includes NGC~3556.

After the initial amplitude and phase calibration, time deviations due to clock drifts, changes in the ionosphere were calibrated. These first steps were performed using \textsc{prefactor} \citep{degasperin_18a}.\footnote{Available and described at \url{https://github.com/lofar-astron/prefactor}}. After that, an initial sky model was constructed using \textsc{PyBDSF} \citep{pybdsf2015}.\footnote{Available at \url{https://github.com/lofar-astron/PyBDSF}}

There are two major data calibration techniques for LOFAR data. 
Both are designed as pipelines but use a different approach to data reduction. First, \textsc{FACTOR}, splits the entire field up in different facets using a Voronoi tessellation \citep{okabe2000} and does a traditional self-calibration on each facet separately, effectively employing a direction-dependent self-calibration.\footnote{Available at \url{https://github.com/lofar-astron/factor}} Individual phase solutions of the facets are then combined to allow for the imaging of the large field of view \citep{weeren2016,williams2016}. The second method, \textsc{KillMS}, applies Kalman filtering and solves for every term in the radio interferometric measurement equation \citep{tasse2014}.

The LoTSS data were calibrated and imaged using \textsc{KillMS}. \cite{Shimwell2017} found that the resulting flux densities of compact and bright ($> 100$~mJy) sources compare very well to values in the TGSS-ADR1 catalogue (TIFR GMRT Sky Survey Alternative Data Release; \citealt{2017A&A...598A..78I}) 
based on observations with the Giant Metrewave Radio Telescope (GMRT). 
However, at high contrast levels, all bright objects reveal a bowl-like artefact. In addition, the calibration scheme employed by \textsc{KillMS} is insensitive to some faint diffuse emission, which might result in up to 20\% underestimated flux densities. These issues are currently rectified for the LoTSS data release 2 (DR2; Shimwell, T. 2018, private communication), but these data were not available yet for NGC\,3556.

\subsection{CHANG-ES}
The CHANG-ES project used the newly upgraded Karl G. Jansky Very Large array (VLA) to observe a sample of 35 galaxies. The science cases of the project included the analysis of the disc-halo interface and star formation properties in general \citep{irwin2012}. The sample was chosen by the following criteria. The inclination of the galaxies had to be larger than $75\degr$ in order to be able to distinguish the disc from the halo. The size of the galaxies had to be between $4\arcmin$, owing to the desired spatial resolution, and $25\arcmin$, owing the field of view of the VLA in the desired frequency and array configuration. 

The observations were conducted in two bands: $1.5$~GHz ($L$ band) and 6~GHz ($C$ band). The different array configurations that were used, B, C, and D, accommodate different resolutions, ranging from about $1\arcsec$ 
in the B array/$C$ band, to about $46\arcsec$ 
in the D-array/$L$-band configuration. The calibration of the CHANG-ES data was conducted with the Common Astronomy Software Application Package \citep[{\small CASA};][]{McMullin2007} and a standard calibration technique was used\footnote{Available at \url{http://casa.nrao.edu}}. Because of the available resolution of the LoTSS data, $\approx$5\arcsec and $\approx$20\arcsec, and the aim of looking at the halo of the galaxy, the $C$ band and $L$ band CHANG-ES data were used for the analysis presented in this work.

\subsection{H$\alpha$ imaging}

We retrieved H$\alpha$ and $R$-band images and the corresponding flat-field and  bias
images from the Isaac Newton Group (ING) archive. The data  were taken on 5 Feburary 2008 with the
Isaac Newton $2.5$ m telescope (INT) and the Wide Field Prime Focus CCD mosaic camera (WFC). Three exposures of 250~s through the  WFCH6568 H$\alpha$ filter (central wavelength
$656.8$~nm, filter band  Full Width Half Maximum (FWHM) $\rm = 9.5~nm$) and one 200 s exposure in the
Harris $R$-band filter  (central wavelength $\rm 638.0~nm$, filter band $\rm FWHM = 152.0~nm$)
were taken. Since NGC\,3556 fits fully into one CCD chip, we only reduced this
part of  the data using the standard CCD reductions tools in
IRAF\footnote{IRAF is distributed by the National Optical Astronomy
Observatory, which is operated by the Association of Universities for
Research in Astronomy (AURA) under a cooperative agreement with the National
Science Foundation.}. 
Pixels that were affected by cosmic rays have been corrected using the Image Reduction and Analysis Facility (IRAF) version of the 
{\small LACOSMIC} routine \citep{vanDokkum2001}. We then shifted the images to
a common reference  frame using 15 stars present on all four images using {\small IMEXA},
{\small IMCENTROID}, and  {\small IMSHIFT} tasks in {\small IRAF} and co-added the three H$\alpha$ images. We
then subtracted the scaled $R$-band image from the combined H$\alpha$ image to
produce a continuum corrected pure emission line image. This image contains mostly the flux of the H$\alpha$ line with a small contribution of [\ion{N}{II}], which has two emission lines also covered by the narrow-band filter used.
A typical value for the ratio of the stronger [\ion{N}{II}] line at $\lambda$ =
$658.3$~nm
to H$\alpha$ is in the range of $0.3$ to $0.4$  in integrated spectra
of star-forming galaxies \citep[e.g.][]{Lehnert1994}, and therefore
somewhat higher
than in \ion{H}{II} regions because of the presence of diffuse ionised gas (DIG)  in and
above the galaxies. 
This makes the contribution of [\ion{N}{II}] emission to our line image
uncertain, but should be
of the order of 40$\%$ for the filter used  in view of the visible 
emission from the diffuse ionised medium \citep[see also][]{Collins2000}.
This contribution is different in \ion{H}{ii} regions and the DIG 
\citep[e.g.][]{Dettmar1992}.  The use of a  broad-band filter as continuum
filter is less exact than the use of a dedicated narrow-band continuum
filter, but is a standard method yielding good results
\citep[e.g.][]{Gildepaz2003}. Since we are only interested in relative fluxes
and especially the morphology of the  thermal gas, we did not perform a
detailed flux calibration. Following basic reduction the $R$-band image was 
astrometrically calibrated using the {\small ASTROMETRY.NET} \citep{lang2010} routines 
and the astrometric solution transferred to the H$\alpha$ image and the 
continuum-corrected H$\alpha$ image.   
Finally we measured the seeing on the H$\alpha$
and $R$-band images by fitting Gauss and Moffat functions to several bright, non-saturated stars. The resulting angular resolution of our set of images is $1\farcs5$.

\begin{table}[!htbp]
\caption{Characteristics of the maps used for the main part of the analysis.}
\label{tableData}
\centering

\begin{tabular}{cccc} 
\hline\hline

Survey  & Frequency     & Sensitivity & Resolution\cr
                & (GHz)         & ($\mu\rm Jy\,beam^{-1}$)  & ($\arcsec$)\cr

\hline

LoTSS           & $0.144$ &     400     &21 \cr 
CHANG-ES$^{a}$          & $1.5$ &       57      &21 \cr 
CHANG-ES$^{b}$          &  6    &       22      &21     \cr 
\hline
\end{tabular}
 \tablefoot{
 $^a$ Combined VLA C and D arrays.\\
 $^b$ VLA D array. 
 } 
\end{table}

\subsection{Imaging}
In Sect.~\ref{sec:results} we present total power and spectral index maps from the CHANG-ES and LOFAR data and a map of the magnetic field strength. For the CHANG-ES data, the image deconvolution with the {\small CLEAN} algorithm was performed via {\small CASA}. For deconvolution, the robust value of the Briggs weighting scheme \citep{Briggs1995} was set to 2 (close to natural weighting) to emphasise faint halo emission. Subsequent images were smoothed, if not indicated otherwise, to a circular synthesised Gaussian beam of $21\arcsec$ FWHM in order to have a uniform set of images while emphasising faint radio haloes. Table~\ref{tableData} summarises the  properties of the images.

\section{Results}
\label{sec:results}
This section is organised as follows: 
In Sect.~\ref{sectResultsTotalPower}, we present the total power maps from LOFAR and CHANG-ES; 
in Sect.~\ref{sectResultsSpectralIndex}, we calculate the radio spectral index for the whole galaxy from the LOFAR and CHANG-ES flux density measurements and published values at other frequencies, and then we present spectral index maps from the CHANG-ES $1.5$ and 6 GHz maps and from the LOFAR 144 MHz and the CHANG-ES $1.5$ GHz images; 
in Sect.~\ref{sectResultsMF},
we present a map of the magnetic field strength derived from the LOFAR 144 MHz image and the spectral index between 144~MHz and $1.5$~GHz and assuming equipartition; 
in Sect.~\ref{sectResultsPI}, CHANG-ES polarisation images and a map of the Faraday rotation measured are presented; 
in Sect.~\ref{subsec:scale_heights}, scale heights are derived from fits to the total power maps and the map of the equipartition magnetic field; 
in Sect.~\ref{sectCRprop}, the data are used to investigate the transport of cosmic rays from the disc into the halo of NGC\,3556;  and
finally, in Sect.~\ref{sectResultsSmallScaleStruct}, we 
examine small-scale structures in the non-thermal and thermal surface brightness distributions traced by  
the high-resolution ($6\arcsec$) LoTSS image and a continuum-subtracted H$\alpha$ image of NGC\,3556.

\begin{figure}[!htbp] 
\includegraphics[width=\columnwidth]{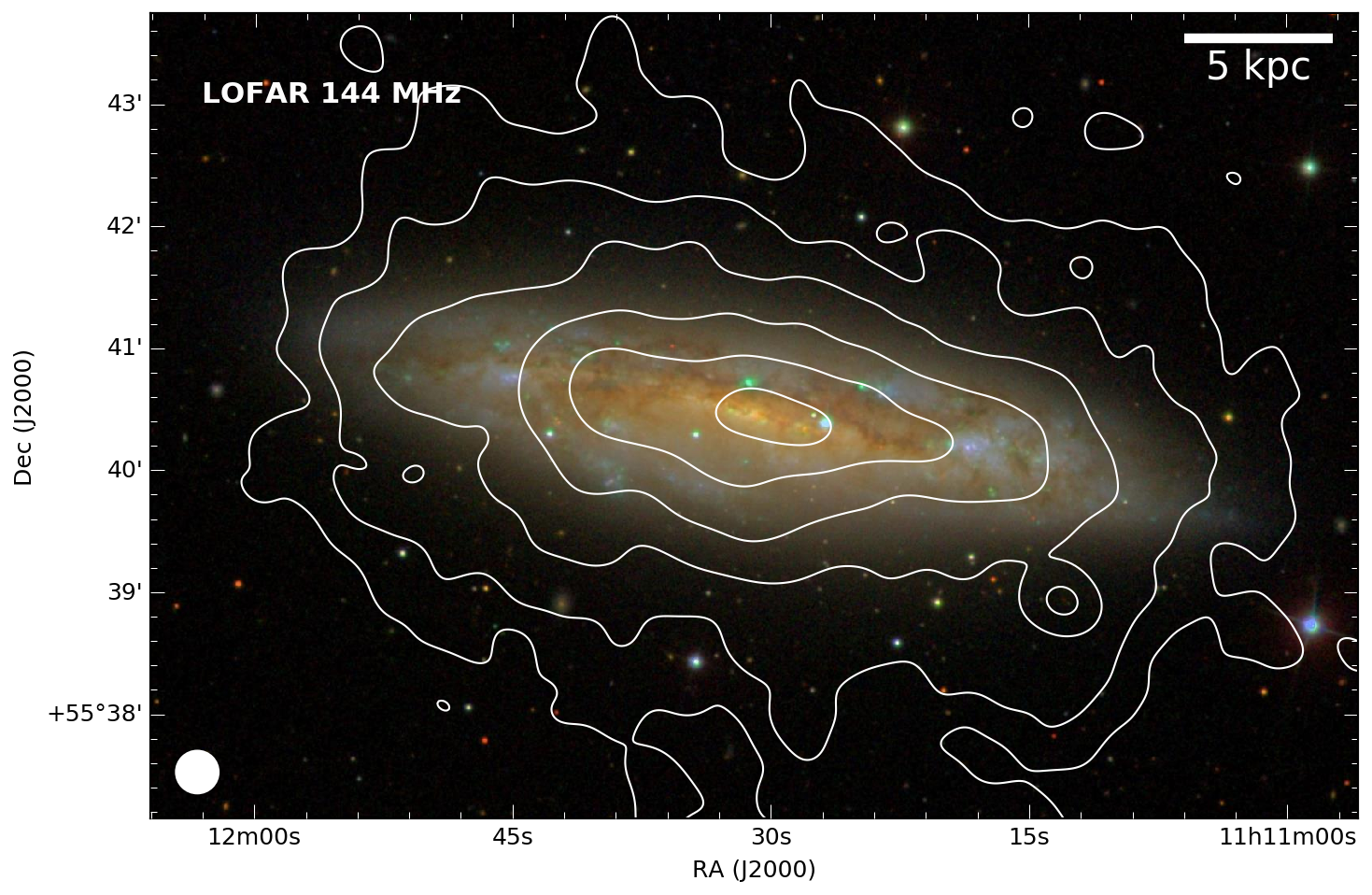}
\includegraphics[width=\columnwidth]{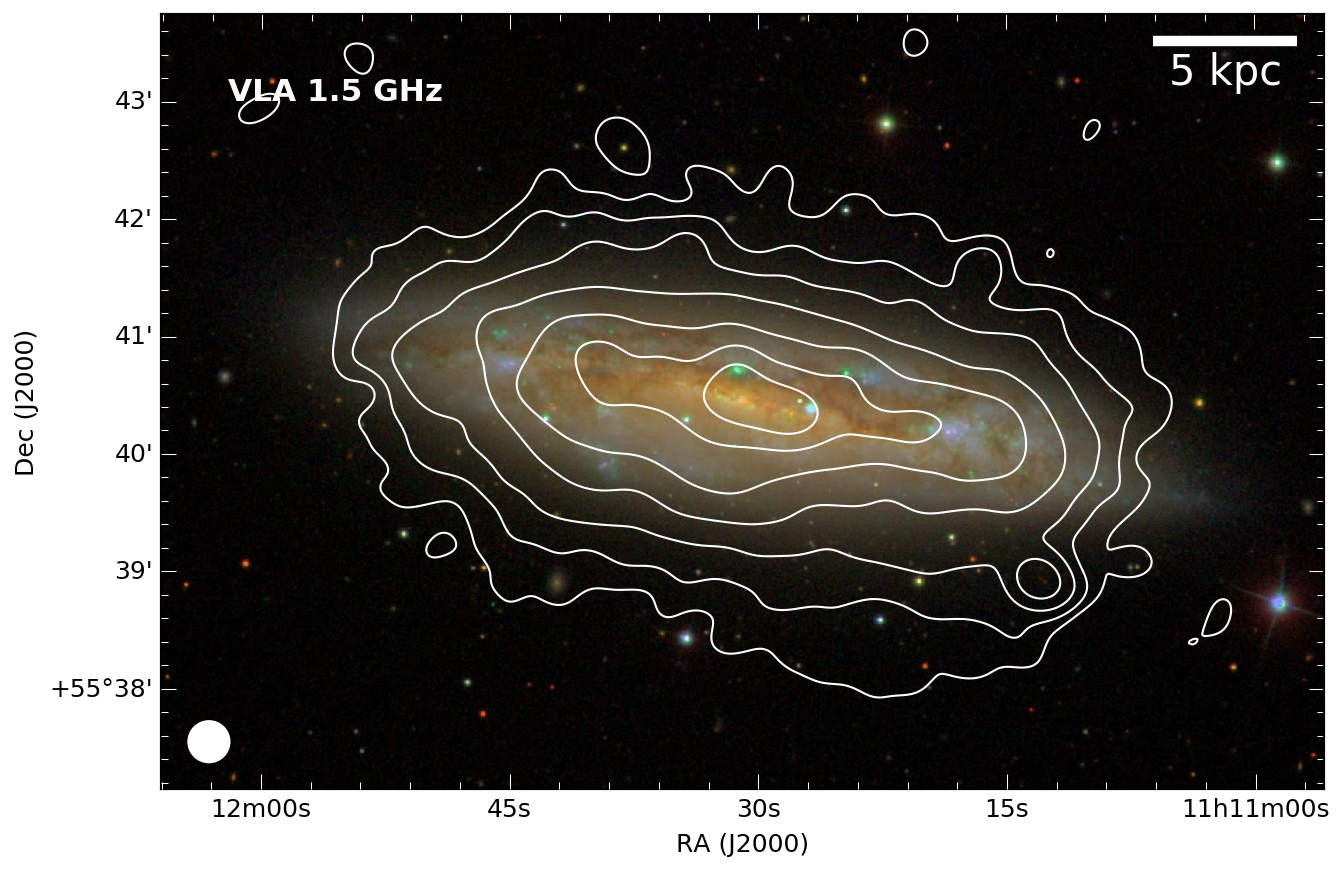}
\includegraphics[width=\columnwidth]{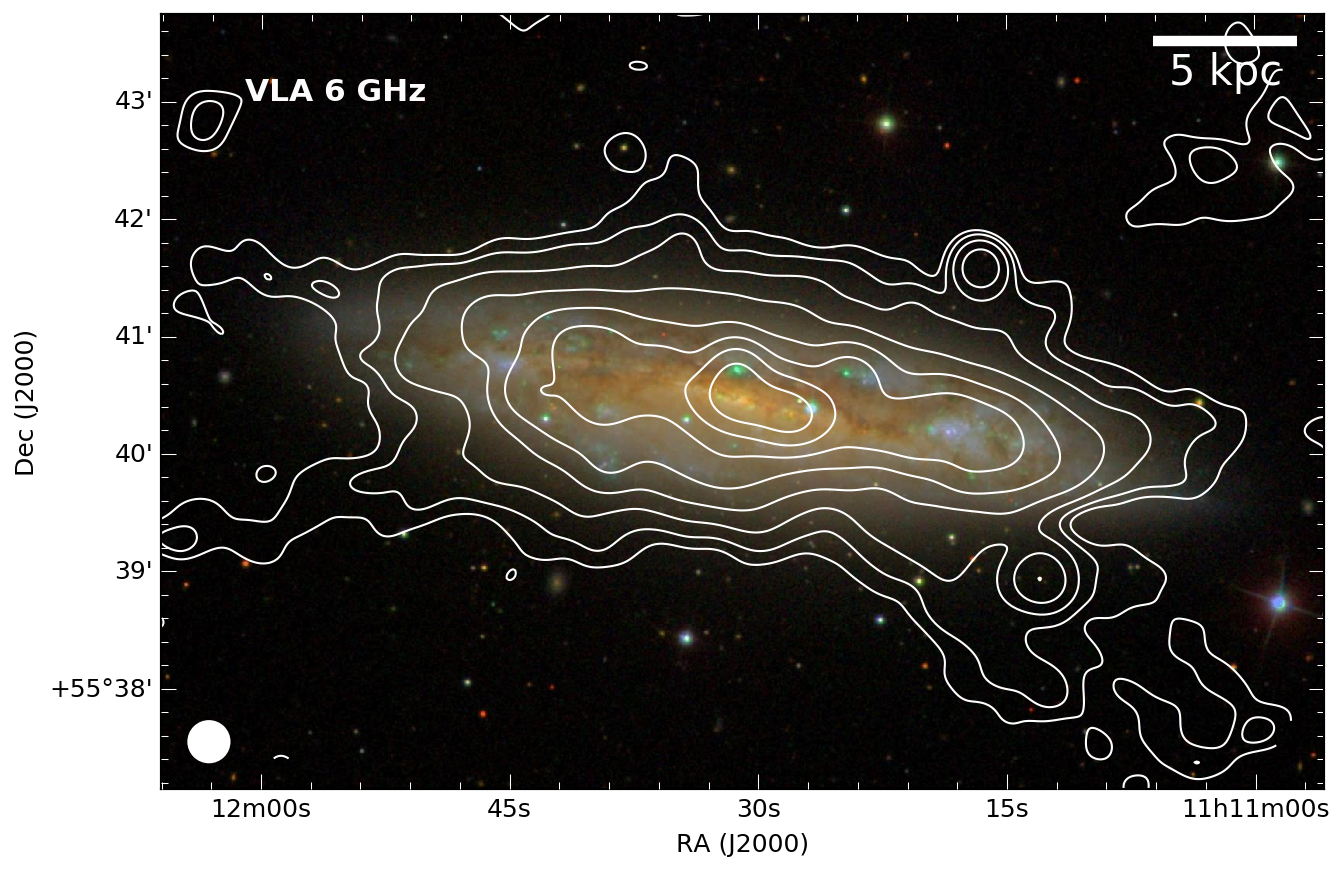}
\caption{Radio images in contours superimposed on a three-colour optical image from SDSS. 
{\it Top panel:} LOFAR 144-MHz image in contours ranging from 3$\sigma = 1.2~\rm mJy\,beam^{-1}$ and increasing by a factor of two.
{\it Middle panel:} 1--2-GHz ($L$ band, combined C and D array) image in contours starting at 3$\sigma = 171~\mu\rm Jy\,beam^{-1}$ and increasing by a factor of two.
{\it Bottom panel:} 4--8-GHz ($C$ band, D array) image in contours starting at 3$\sigma = 66~\mu\rm Jy\,beam^{-1}$ and increasing by a factor of two. Some {\scriptsize CLEAN} artefacts are visible in the form of symmetric features on the south side of the galaxy.
In all panels, the synthesised beam of $21\arcsec$ FWHM is shown as a filled white circle in the bottom left corner. 
} 
\label{figsLOFARVLA}
\end{figure}

\subsection{Total power}
\label{sectResultsTotalPower}

In Fig.~\ref{figsLOFARVLA} (top panel), we show the LOFAR 144 MHz image in contours superimposed on a three-colour optical image from SDSS DR14 \citep{sdss_dr14}. The radio synchrotron halo extends up to 10~kpc into the halo, while the radial extent is comparable to the extent of the galaxy at optical wavelengths. The total flux density measured within the first contour level amounts to $1.43 \pm 0.36$~Jy. This is in agreement with the 151 MHz measurement from the 6C survey \citep{6c_p3} within the uncertainties. 

In the middle and bottom panels of Fig.~\ref{figsLOFARVLA}, we show the CHANG-ES VLA images in contours superimposed on the same optical image from the SDSS. The $1.5$ GHz image ($L$ band) is presented in the middle panel 
and the 6 GHz image ($C$ band, D array) is presented in the bottom panel. The higher frequency CHANG-ES maps show a lesser vertical extent of the halo than the LOFAR map, although the extent of the total intensity along the major axis barely changes.

In $L$ band, data from the C and D arrays have been combined to attain the extended emission found in compact arrays, while keeping a resolution close to that of the extended array. Within the first contour, the total power flux density at $1.5$~GHz amounts to $280 \pm 10$~mJy, where the first contour starts at a noise level of 3$\sigma$ ($171~\mu\rm Jy\,beam^{-1}$). In $C$ band (6~GHz), the total flux density measured within the first contour amounts to $92 \pm 3$~mJy, where the first contour starts at a level of 3$\sigma$ ($66~\mu \rm Jy\,beam^{-1}$). At $1.5$~GHz, the flux density agrees within the uncertainties with that of \citet{changes_dr1}, who used only the D array data. At 6~GHz, our flux density is slightly higher than that of \citet{changes_dr1}, which we ascribe to our lower angular resolution, helping us to detect faint emission in the halo. 

\begin{figure}[!thbp]
\includegraphics[width=\columnwidth]{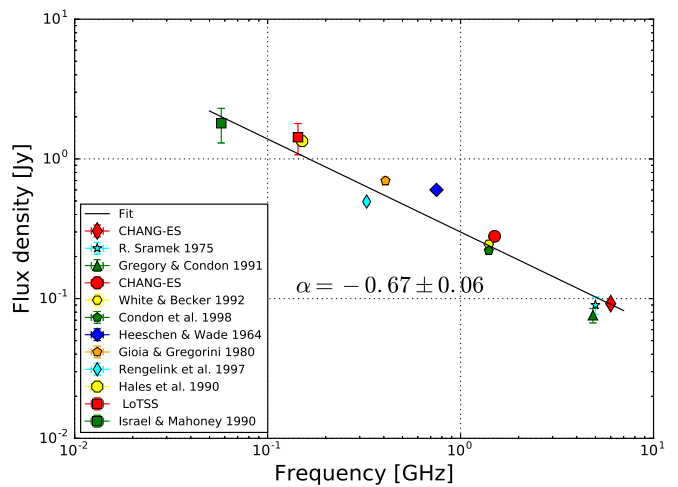}
\caption{Flux density measurements of NGC\,3556 from the published literature and from this work. A linear fit to the logarithmic values results in a spectral index of $-0.67 \pm 0.06$.}
\label{allfluxes}
\end{figure}

\begin{figure}[!htbp]
\includegraphics[width=\columnwidth]{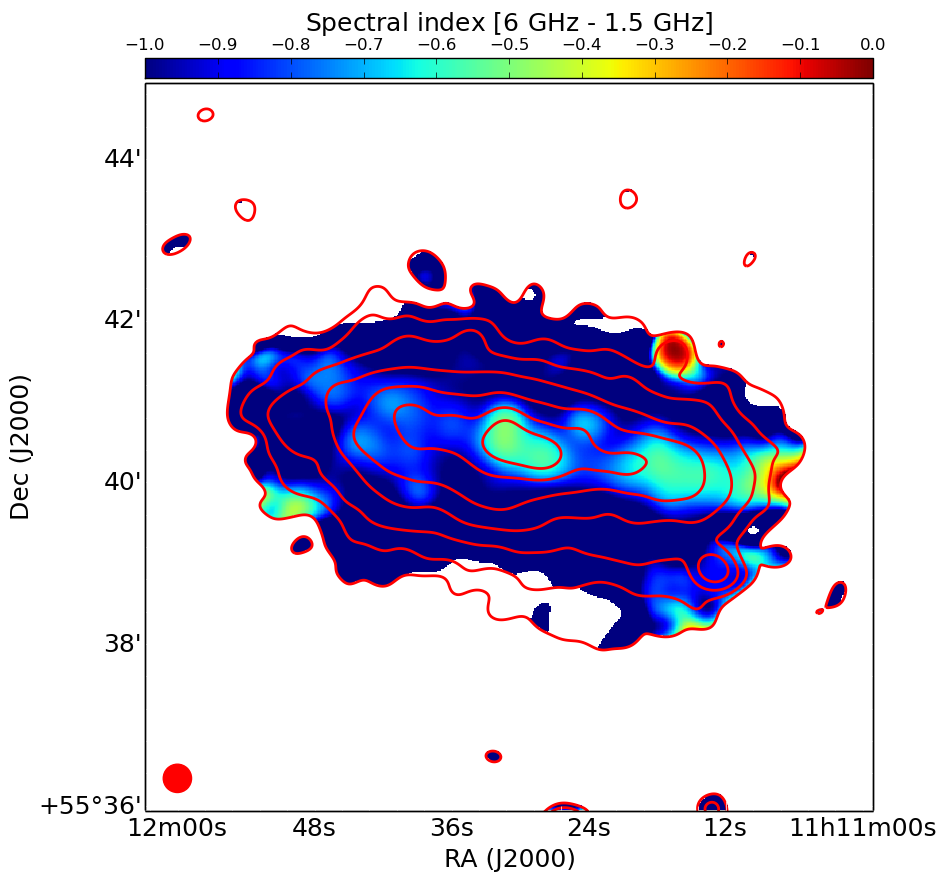}
\includegraphics[width=\columnwidth]{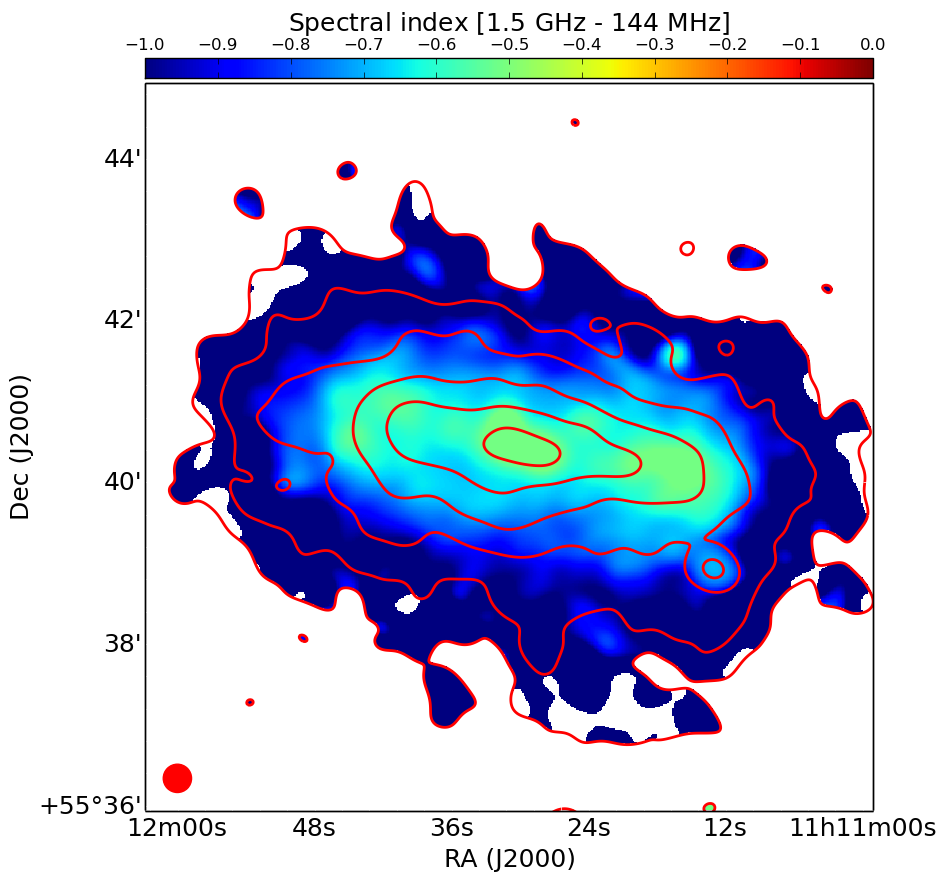}
\caption{Radio spectral index maps. 
{\it Top panel:} From the CHANG-ES data calculated between 
$1.5$ and 6~GHz.
{\it Bottom panel:} Between the 144 MHz LOFAR and the $1.5$ GHz VLA maps.
Red contours show the surface brightness distribution of the $1.5$ GHz map (top) and 144 MHz map (bottom) with the same contour levels as in  Fig.~\ref{figsLOFARVLA}. 
} 
\label{figSPIX}
\end{figure}

\subsection{Spectral index}
\label{sectResultsSpectralIndex}

Comparing the obtained flux densities to known values from the literature at different frequencies allows us to calculate the global radio continuum spectrum (\cite{Sramek1975},\cite{Gregory1991},  \cite{White1992}, \cite{Condon1998}, \cite{Heeschen1964}, \cite{Gioia1980}, \cite{Rengelink1997}, \cite{Hales1990}, \cite{Israel1990}). In Fig.~\ref{allfluxes}, we show a logarithmic 
plot of the literature values and a fit to the values, resulting in a spectral index of $\alpha = -0.67\pm 0.06$. This fits well with the expected synchrotron spectral index of $\alpha=-0.7$ as typically found in spiral galaxies \citep{Beck2015}.

We also looked at the spatially resolved spectral index distribution, both between $C$ and $L$ band and between $L$ band and LOFAR 144~MHz. In Fig.~\ref{figSPIX}, we show both spectral index maps.

The spectral index map between $C$ and $L$ band shows a thin disc with a spectral index of about $-0.7$, except for the central and western star-forming regions where the spectral index is $\approx$-0.5. 
In the spectral index map between $144$~MHz and $L$ band, 
no clear disc structure is seen and the disc and halo appear to have a very similar spectral index ($\alpha \approx -0.5$). At a distance of about $8$~kpc from the disc, the spectrum steepens to values of $\alpha \approx -1$ and lower. The steepening in the halo is expected if spectral ageing plays an important role in the halo.

\subsection{Magnetic field}
\label{sectResultsMF}

We calculated magnetic field maps from the total intensity maps using the equipartition assumption, which states that the energy densities of cosmic rays and magnetic fields are equal in galaxies. Following \cite{Beck2005}, the magnetic field strength is calculated using
\begin{equation}
\label{beckeq}
B_{\rm eq} = \dfrac{4\pi (1-2\alpha) (K_0 + 1) I_\nu E_p^{1+2\alpha} 
(\nu / 2c_1)^{-\alpha}}{[(-2\alpha -1) c_2(-\alpha) l c_4(i)]^{1/(3 - \alpha)}} \,
,\end{equation}
where $\alpha$ is the synchrotron spectral index, defined as $I_\nu \propto \nu^\alpha$,  where \textit{I$_\nu$} is the synchrotron intensity at frequency $\nu$. The value $K_0$ is a constant factor representing the ratio of the proton to electron number density. This constant is usually assumed to be $\approx$100. This value, however, increases towards the halo of galaxy, which makes the derived strengths lower limits \citep{Beck2005}. \cite{Lacki2013} also show that $K_0$ remains the same to an order of magnitude even in starburst galaxies such as M\,82. Furthermore, \textit{l} represents the path length along the line of sight through the galaxy, assumed to be $20$~kpc, \textit{i} is the inclination angle of the galaxy, $E_p$ is the proton rest mass, and \textit{c$_{1-4}$} are constants. 

To produce a map of the equipartition magnetic field strength, we used the $144$~MHz map and the spectral index map
between $144$~MHz and $1.5$~GHz (bottom panel of Fig.~\ref{figSPIX}), but blanked pixels with $\alpha > -0.5$ because such regions are likely contaminated by free-free absorption, which would result in significantly underestimated values of the magnetic field strength.

\begin{figure}[!htbp]
\includegraphics[width=\columnwidth]{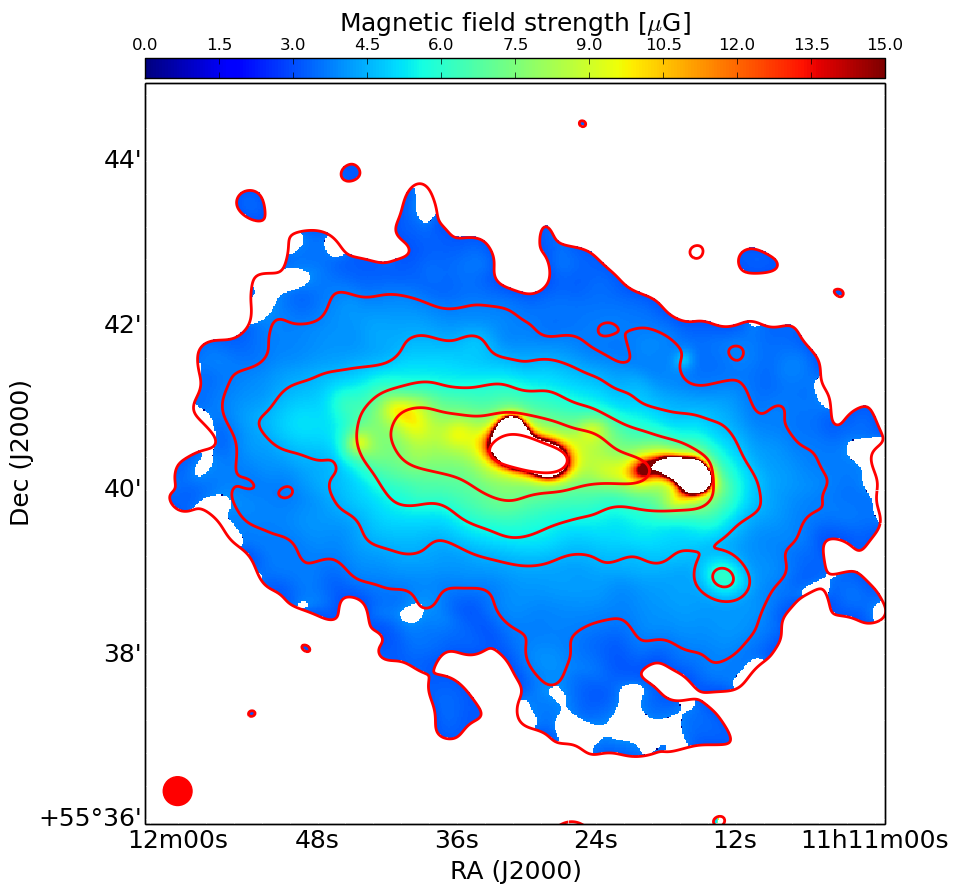}
\caption{Magnetic field strength map derived through the equipartition assumption using the spectral index between $L$ band and $144$~MHz and synchrotron flux density at $144$~MHz. Red contours show the surface brightness distribution of the $144$~MHz map as shown in Fig.~\ref{figsLOFARVLA}. 
The central region and a region to the west (both in white) were blanked because the spectral index $\alpha$ is higher than $-0.5$.
}
\label{mag_L}
\end{figure}

The resulting magnetic field strength map is presented in Fig.~\ref{mag_L}.
The galaxy exhibits magnetic field strengths around 10 
$\mu$G within the disc and smaller field strengths around 5~$\mu$G in the halo. This is in the same range as found in other spiral galaxies (e.g. \citealt{Beck2015}). The average uncertainty of the estimates of the magnetic field strength is about 10\%.

\subsection{Polarised intensity}
\label{sectResultsPI}

\begin{figure}[!htbp]
\includegraphics[width=\columnwidth]{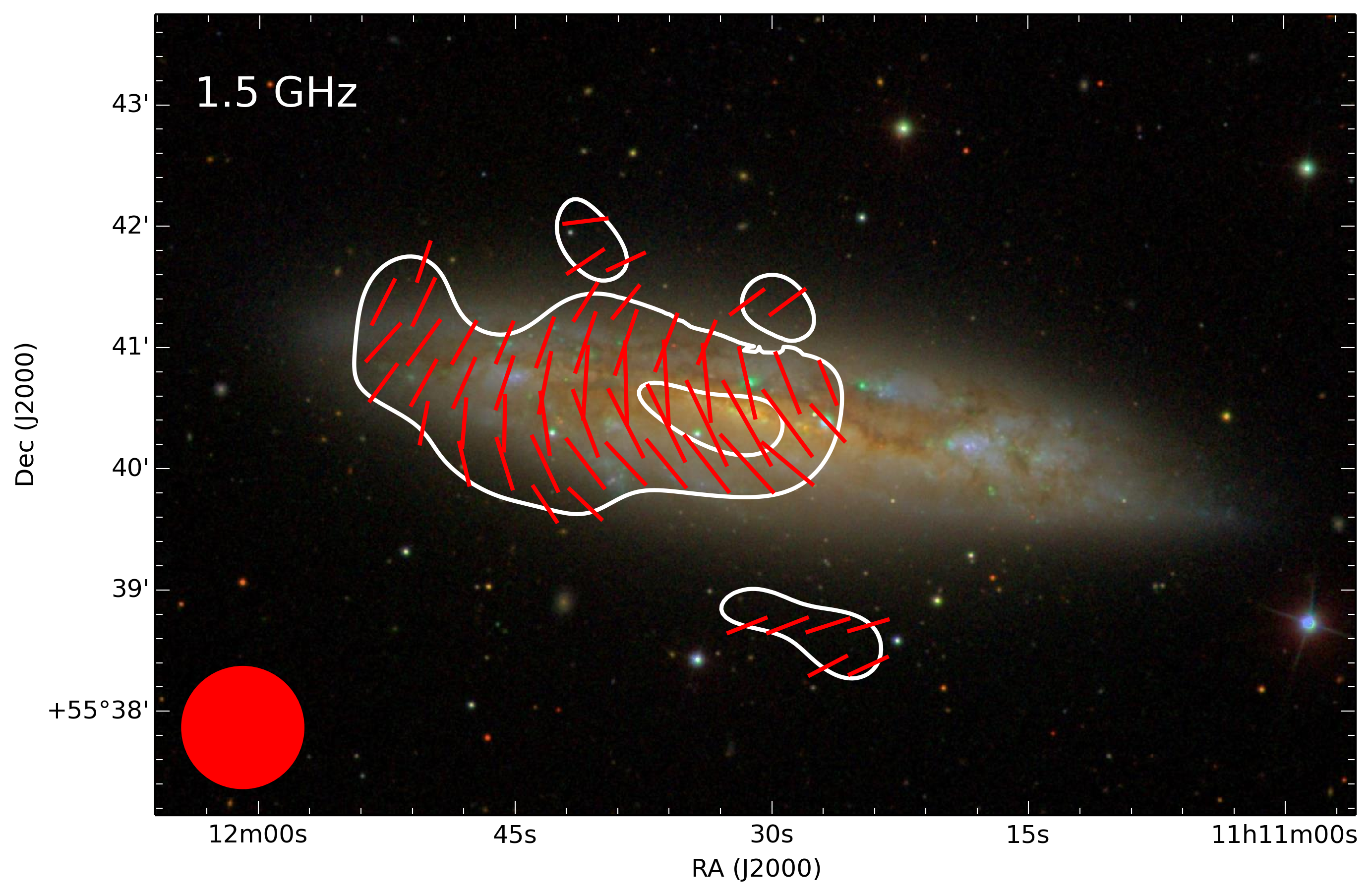} 
\includegraphics[width=\columnwidth]{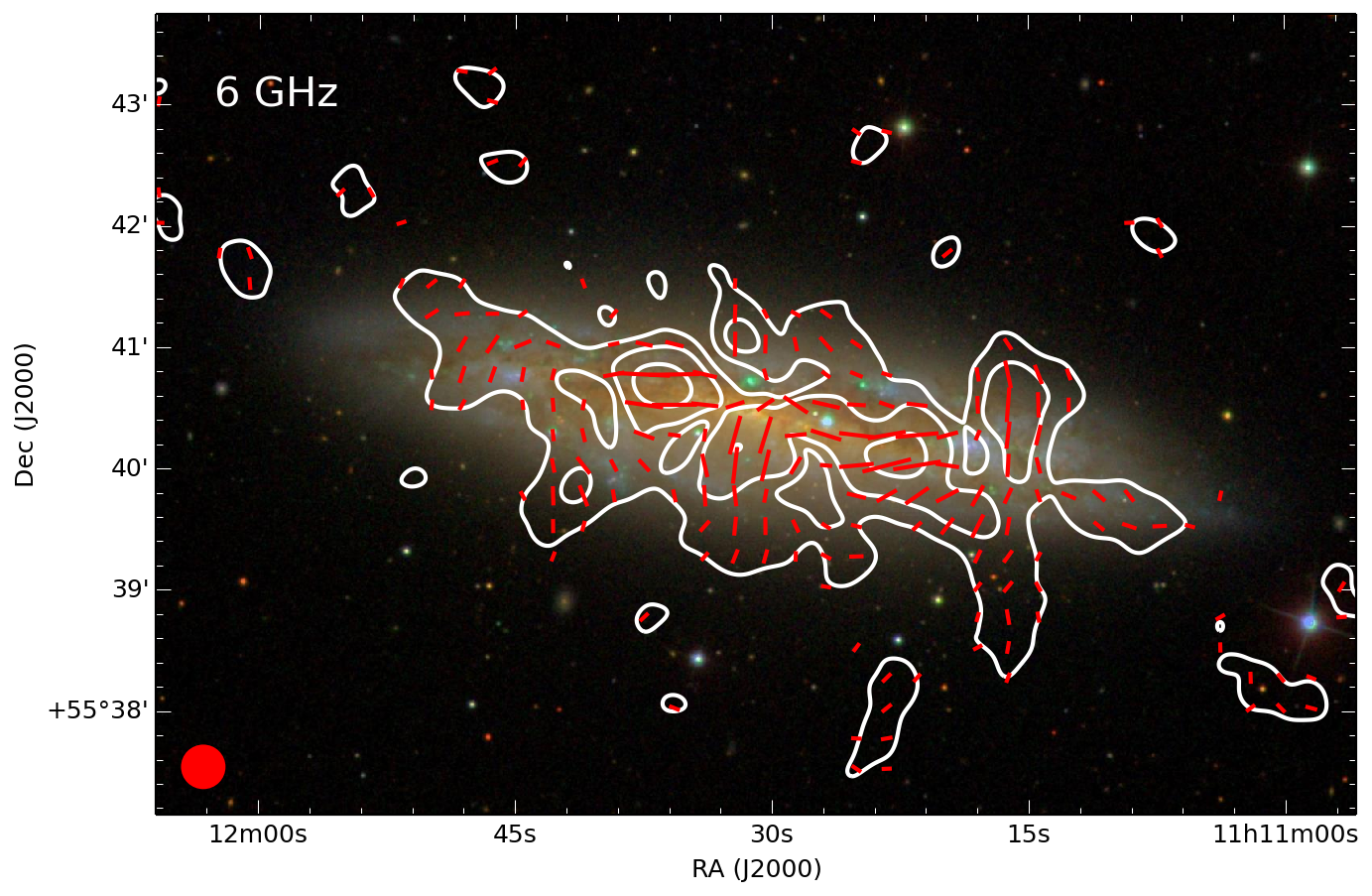} 
\includegraphics[width=\columnwidth]{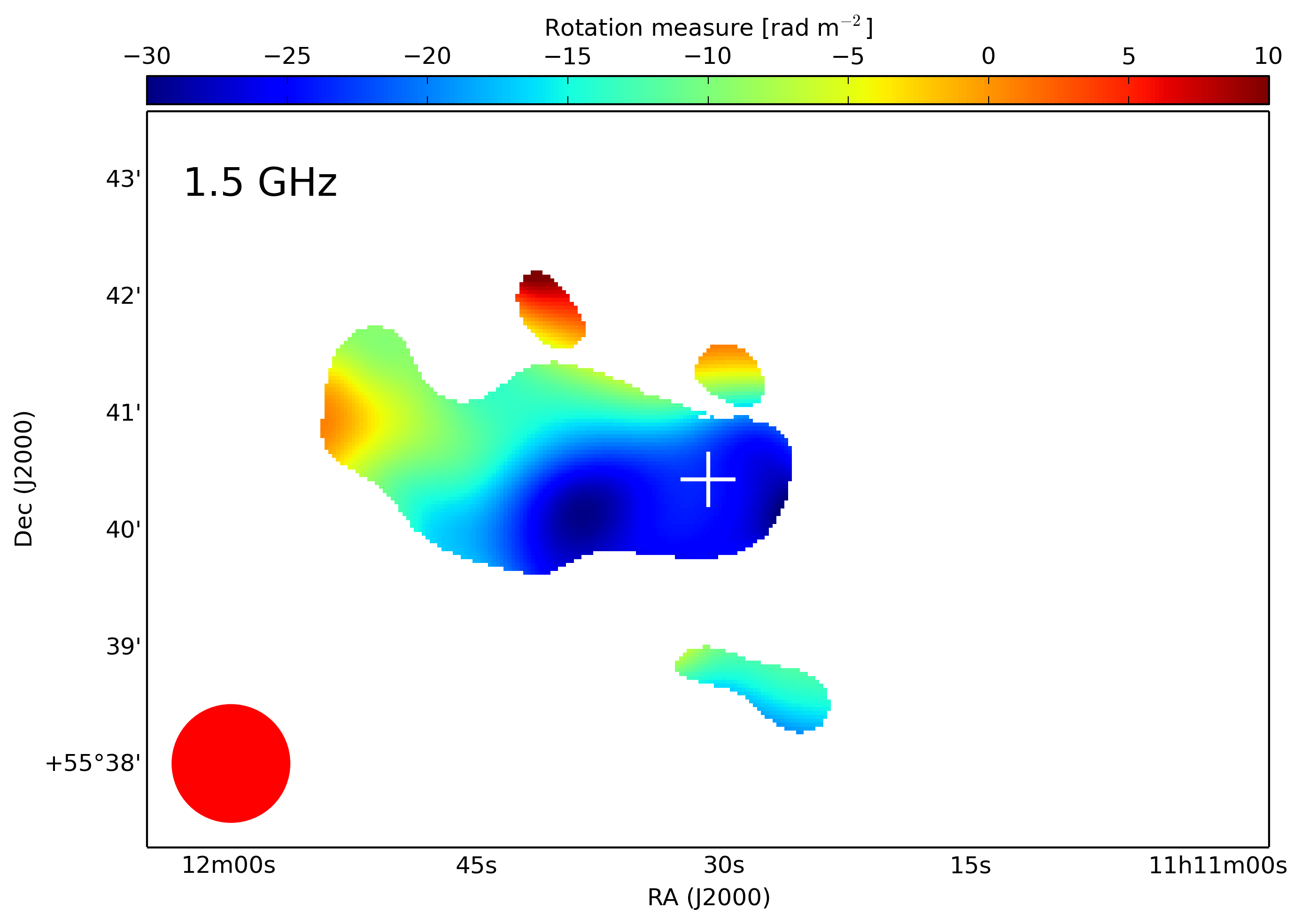} 
\caption{
{\it Top panel:} Magnetic field vectors (red) and PI contours from the CHANG-ES D-array/$L$-band data overlaid on the same optical image as in Fig.~\ref{figsLOFARVLA} at $57\arcsec$ FWHM resolution. 
Contours start at $3\sigma = 90~\mu\rm Jy\, beam^{-1}$, increasing by a factor of two, and red lines indicate magnetic field vectors.
{\it Middle panel:} Magnetic field vectors (red) and PI contours from the D-array/$C$-band CHANG-ES data overlaid onto the image as in Fig.~\ref{figsLOFARVLA} at FWHM 21\arcsec resolution. Contours start at a PI level of $3\sigma=27~\mu\rm Jy\,beam^{-1}$, increasing by a factor of two.
{\it Bottom panel:} Rotation measure map from the CHANG-ES D-array/$L$-band data with values within the 3$\sigma$ contour of the PI map from the $L$-band data (top panel). 
The white cross indicates the position of the centre of the galaxy. Filled red circles indicate the synthesised beam size.
}

\label{figPIBRM}
\end{figure}

CHANG-ES data provides us with full Stokes information, which we used in combination with rotation measure synthesis \citep[RM synthesis;][]{brentjens2005} to obtain the polarised intensity (PI) and the orientation of the ordered magnetic field component present in NGC\,3556. To see the extent of the magnetic field we used the D-array/$L$-band data because of its sensitivity to large angular scales. 
In Fig.~\ref{figPIBRM}, we show magnetic fields vectors (in red) in regions where the PI in the $L$-band CHANG-ES image is larger than 3$\sigma$ ($90~\mu\rm Jy\,beam^{-1}$) overlaid on the SDSS optical image. The contours show the PI, starting at the  3$\sigma$ level. The polarised emission at $1.5$~GHz is limited to the eastern side of the galaxy. The total polarised flux density is $1.2 \pm 0.2$~mJy.

The middle panel of Fig.~\ref{figPIBRM}  
shows the PI map and the corresponding magnetic field vectors from the D-array/$C$-band map. The polarisation is no longer limited to a single patch, but is observed across the whole galaxy.

The large bandwidth combined with the high spectral resolution allows us to study the Faraday rotation measure (RM) of the galaxy. We used the RM synthesis technique developed by \citet{brentjens2005} to obtain the RM map shown in the bottom panel of Fig.~\ref{figPIBRM}. 
The RM map from the D-array/$L$-band data shows values between $-30$ and $10~\rm rad\,m^{-2}$ within the first (3$\sigma$) $1.5$ GHz PI contour (Fig.~\ref{figPIBRM}, top panel), with positive RM values found only outside the visible optical extent of the galaxy as shown by the SDSS image. The foreground RM value of the Milky Way, measured from a map provided by \cite{Oppermann2015}, is approximately $6 \pm 5~\rm rad\,m^{-2}$. 

\subsection{Scale heights}
\label{subsec:scale_heights}

To find the scale heights of the total power profile, exponential functions were fitted to the data. Following \cite{Dumke1995}, an intrinsic exponential profile, 
\begin{equation}
w(z) = w_0\,\textnormal{exp}(-z/z_0) \, ,
\end{equation}
is convolved with the beam of the telescope which, for a map that has been deconvolved with the {\small CLEAN} algorithm, is
a Gaussian profile, 
\begin{equation}
g(z) = \frac{1}{\sqrt{2\pi\sigma^2}}\,\textnormal{exp}(-z^2/2\sigma^2 ) \, ,
\end{equation}
resulting in the expected distribution of the galaxy which is described by 
\begin{multline}
\label{eq:doubleexp}
W_{\textnormal{exp}}(z) = \frac{w_0}{2}\,\textnormal{exp}(-z^2/2\sigma^2)\left[\textnormal{exp}\left( \frac{\sigma^2 - zz_0}{\sqrt{2}{\sigma z_0}} \right) ^2 \,\textnormal{erfc}\left(\frac{\sigma^2-zz_0}{\sqrt{2}\sigma z_0} \right) \right. \\ + \left. \textnormal{exp}\left( \frac{\sigma^2 + zz_0}{\sqrt{2}{\sigma z_0}} \right) ^2 \,\textnormal{erfc}\left(\frac{\sigma^2+zz_0}{\sqrt{2}\sigma z_0} \right)\right]\, ,
\end{multline}
where $w_0$ is the amplitude and $z_0$ the scale height of the distribution, and erfc is the complementary error function
\begin{equation}
\textnormal{erfc}(x) = \frac{2}{\sqrt{\pi}} \int_{x}^{\infty} \textnormal{exp}(-r^2){\rm d}r \, .
\end{equation}

Since most edge-on spiral galaxies show two disc components \citep{thick_2006}, the so-called thin and thick disc, a two-component exponential function is fitted using Equation~(\ref{eq:doubleexp}) twice. These two components are characterised by their amplitudes $w_1$ and $w_2$ and scale heights $z_1$ and $z_2$, which replace variable $w_0$ and $z$ in Eq.~(\ref{eq:doubleexp}). The sum of these two components is then fitted to the inclination-corrected profile. Because of the high inclination of $81\degr$ a simple $\sin(i)$ correction was performed. The quality of each fit is determined by a reduced $\chi^2$ analysis. The profile was computed from box integrations on the galaxy, having 21 boxes across the minor axis, each having a size of $21\arcsec \times 300\arcsec$. The large size across the major axis ensures a global average.

\subsubsection{Total power}
\begin{figure}[!htbp]
\includegraphics[width=\columnwidth]{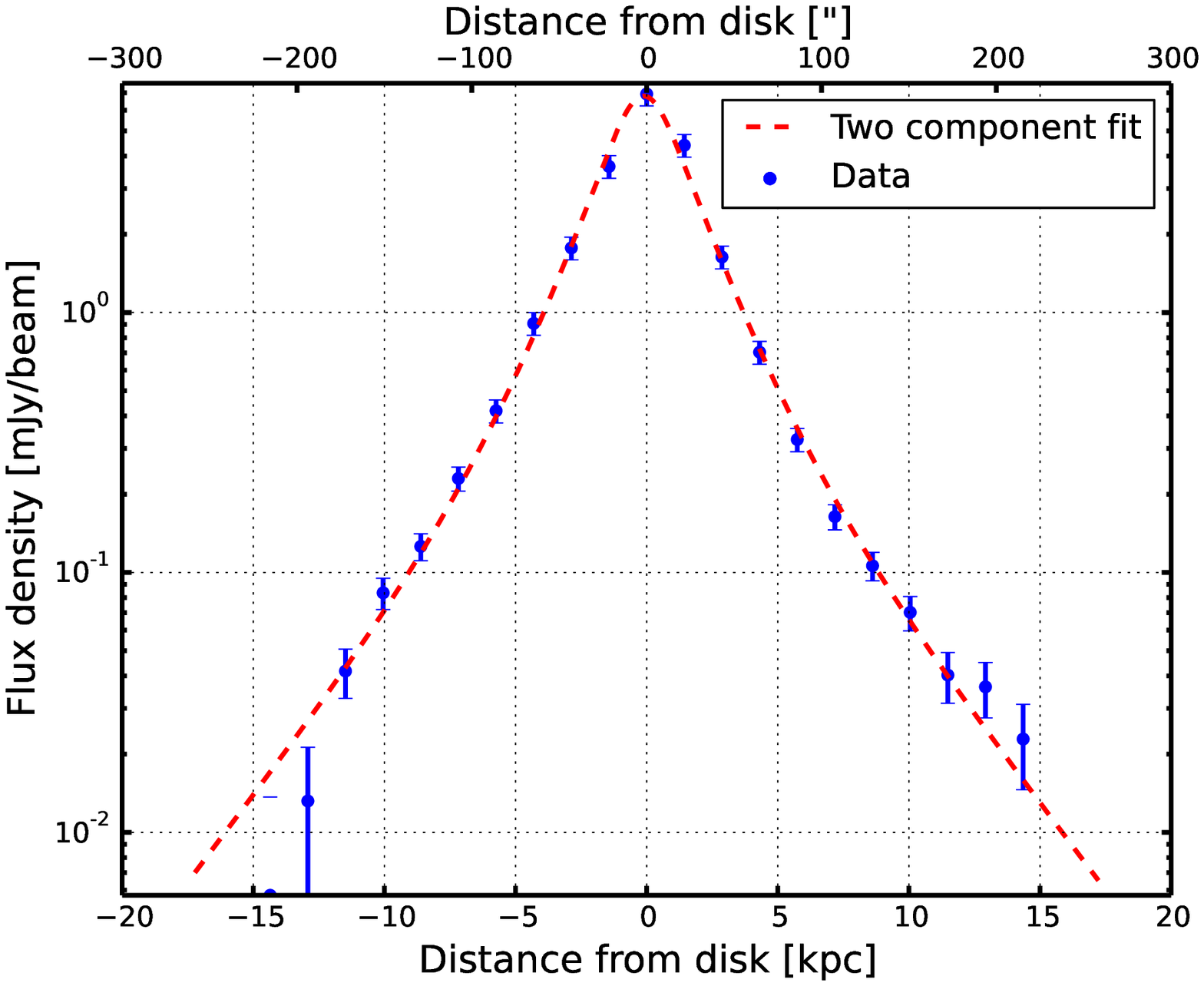}
\includegraphics[width=\columnwidth]{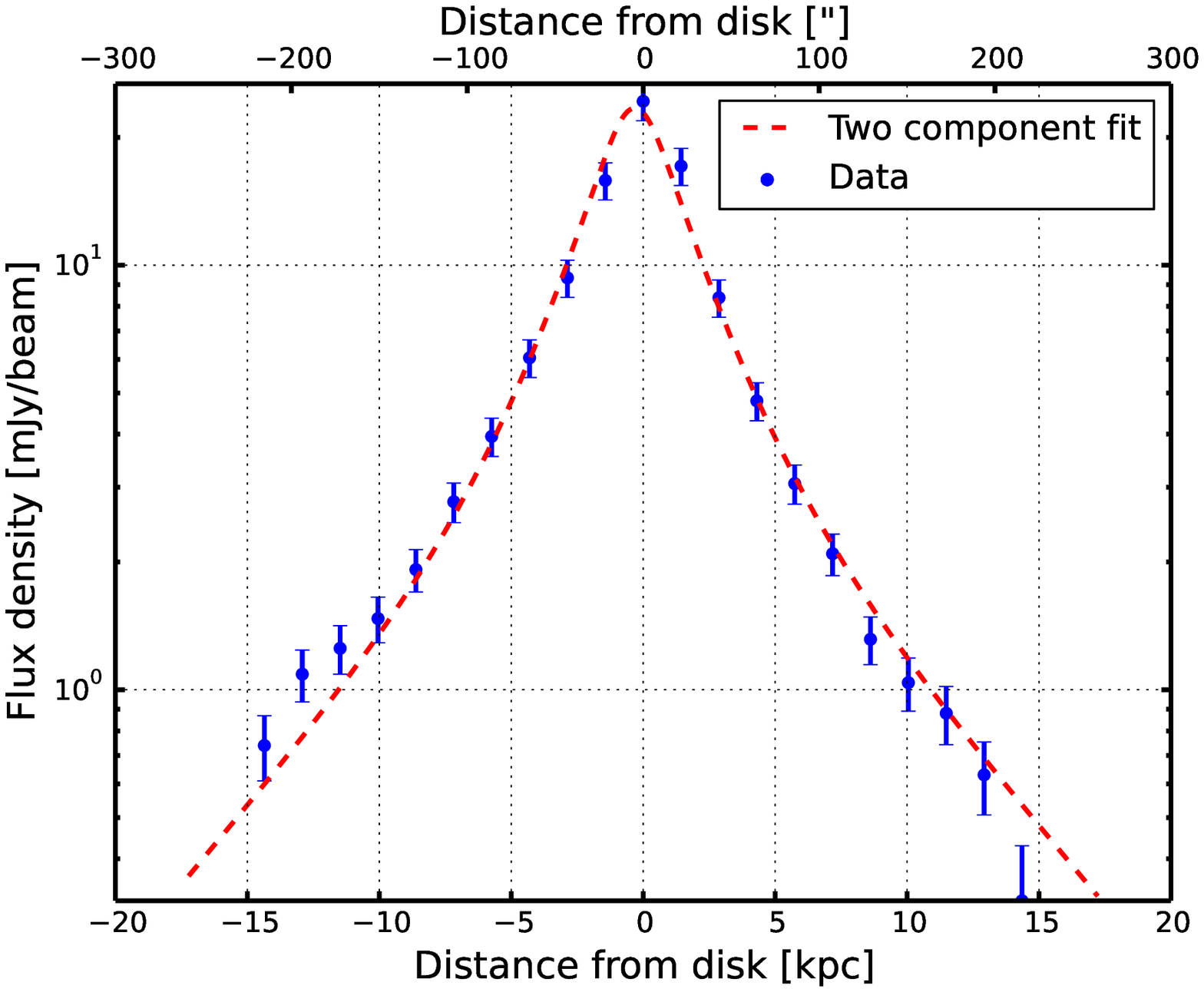}
\caption{Total power profile plots of NGC 3556 derived from the VLA $1.5$ GHz map (top) and LOFAR 144 MHz map (bottom). Blue points represent measured data and red dashed lines represent fits to it.}
\label{scales_TP}
\end{figure}

In Fig.~\ref{scales_TP}, we present the results from the scale height analysis of the $1.5$ GHz and 144 MHz maps with the best-fit parameters listed in Table~\ref{TP_scaleheiths_table}.
The fits are of good quality and reveal 
two distinct components with scale heights that are in good agreement with measurements for other edge-on galaxies \citep[e.g.][]{changes9}. As expected from previous measurements and from the effects of spectral ageing, the extent of the halo of the galaxy is larger at $144$~MHz than at $1.5$~GHz.

\begin{table}[!htbp]
\caption{
Best-fit flux density amplitudes and scale heights of thin and thick discs for the VLA $1.5$ GHz map and the LOFAR 144 MHz map.}
\label{TP_scaleheiths_table}
\centering
\begin{tabular}{ccccc}
\hline\hline 
Parameter & $1.5$~GHz & 144~MHz & Unit\\
\hline
$w_1$ & $7.8 \pm 1$ & $22.6 \pm 3.7$ & $\rm mJy\,beam^{-1}$\\
$z_1$ & $1.4 \pm 0.2$ & $1.9 \pm 0.5$ & kpc\\
\hline
$w_2$ & $1.3 \pm 1.1$ & $6.2 \pm 4.3$ & $\rm mJy\,beam^{-1}$\\
$z_2$ & $3.3 \pm 0.8$ & $5.9 \pm 1.9$ & kpc\\
\hline
$\chi_{\rm red}^ 2$ & $1.23$ & $1.47$ &\\
\hline
\end{tabular}
\end{table}

\subsubsection{Magnetic field}
\begin{figure}[h]
\includegraphics[width=\columnwidth]{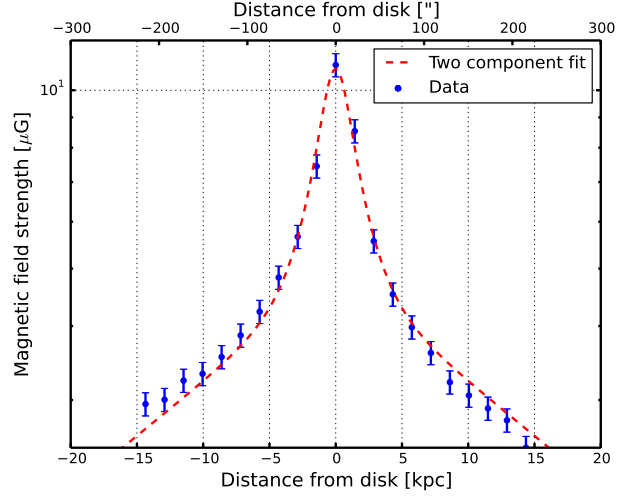}
\caption{Profile of the magnetic field strength in NGC 3556 derived from the LOFAR $144$~MHz map. The blue points represent measured data and the red dashed line the fit to the data.
}
\label{magscaleheight}
\end{figure}

We analysed the magnetic field strength profile in the same way as the profiles of the total power maps. 
The resulting amplitudes and scale heights of the thin and thick disc are listed in Table~\ref{Mag_scaleheiths_table}. 
Figure~\ref{magscaleheight} shows the magnetic field scale height derived from the magnetic field map as derived from LOFAR 144~MHz. As for the radio continuum intensities, the magnetic field profile shows two distinct components.  While the scale height of the thick magnetic field component appears to be very high ($23.7$ $\pm$ $4.1$~kpc), the value is consistent with results by \citet{Pakmor2017} who analysed magnetic field simulations of Milky Way-like galaxies.

\begin{table}[!htbp]
\caption{List of magnetic field strengths and scale heights in the thin and thick discs as derived from the LOFAR $144$~MHz map.
}
\label{Mag_scaleheiths_table}
\centering
\begin{tabular}{ccc}
\hline\hline
Parameter & Value & Unit\\
\hline
$B_1$ & $8.2 \pm 0.9$ & $\mu$G\\
$h_{B1}$ & $1.5 \pm 0.3$ & kpc\\
\hline
$B_2$ & $4.9 \pm 0.4$ & $\mu$G\\
$h_{B2}$ & $23.7 \pm 4.1$ & kpc\\
\hline
$\chi^2_{\rm red}$ & $1.56$ &\\
\hline
\end{tabular}
\end{table}

\subsection{Cosmic-ray propagation}
\label{sectCRprop}
We now use our data to investigate the transport of cosmic rays from the disc into the halo, using the electrons as proxies for GeV protons and heavier nuclei. Our model assumes that the electrons are injected in the galactic midplane at $z=0~\rm kpc$ following a power law with an injection spectral index $\gamma$, so that the CRe number density is $N(E,z=0)\propto E^{-\gamma}$ where $E$ is the CRe energy. The CRe number density is then evolved as function of distance from the disc using equations for pure diffusion and advection \citep{heesen2016}. Energy losses that are taken into account include synchrotron and inverse Compton (IC) radiation and adiabatic losses. The model neglects other types of losses, such as ionisation and bremsstrahlung losses. At frequencies below and around 1~GHz, these types of losses are only important in dense gaseous regions 
in the disc plane (e.g. \citealt{Basu2015}).  

The CRe number density is then convolved in frequency space with the synchrotron emission spectrum of an individual CRe to calculate synchrotron intensities. Finally, this vertical intensity profile is convolved with the effective beam (Section~\ref{subsec:scale_heights}), such that they can directly be compared with the observations. These steps are carried out in the SPectral INdex Numerical Analysis of K(c)osmic-ray Electron Radio-Emission \citep[{\small SPINNAKER;}][]{heesen2016} computer program, for which we now provide a graphical user interface with {\small SPINTERACTIVE}.\footnote{SPINNAKER and SPINTERACTIVE can obtained from: \href{www.github.com/vheesen/Spinnaker}{www.github.com/vheesen/Spinnaker}}

We model the vertical profile of the magnetic field strength as
\begin{equation}
B(z) = B_1\exp\left (-\frac{z}{h_{B1}}\right ) + (B_0 - B_1)\exp\left (-\frac{z}{h_{B2}}\right ),
\end{equation}
where $B_1$ is the magnetic field strength of the thin disc component and $h_{\rm B1}$ and $h_{\rm B2}$ are the magnetic field scale heights of the thin and thick disc components, respectively. The value $B_0$ is the total magnetic field strength in the disc, which we fix as $B_0=9~\mu\rm G$. We simultaneously fit the floating parameters of the magnetic field model together with the diffusion coefficient or advection speed, when either fitting the diffusion or advection model, respectively. The CRe injection index $\gamma$ is another free parameter. The global quality of the fit was determined by calculating the average reduced $\chi^2$ from the values for each profile.

\begin{figure*}[!h]
\includegraphics[width=17cm]{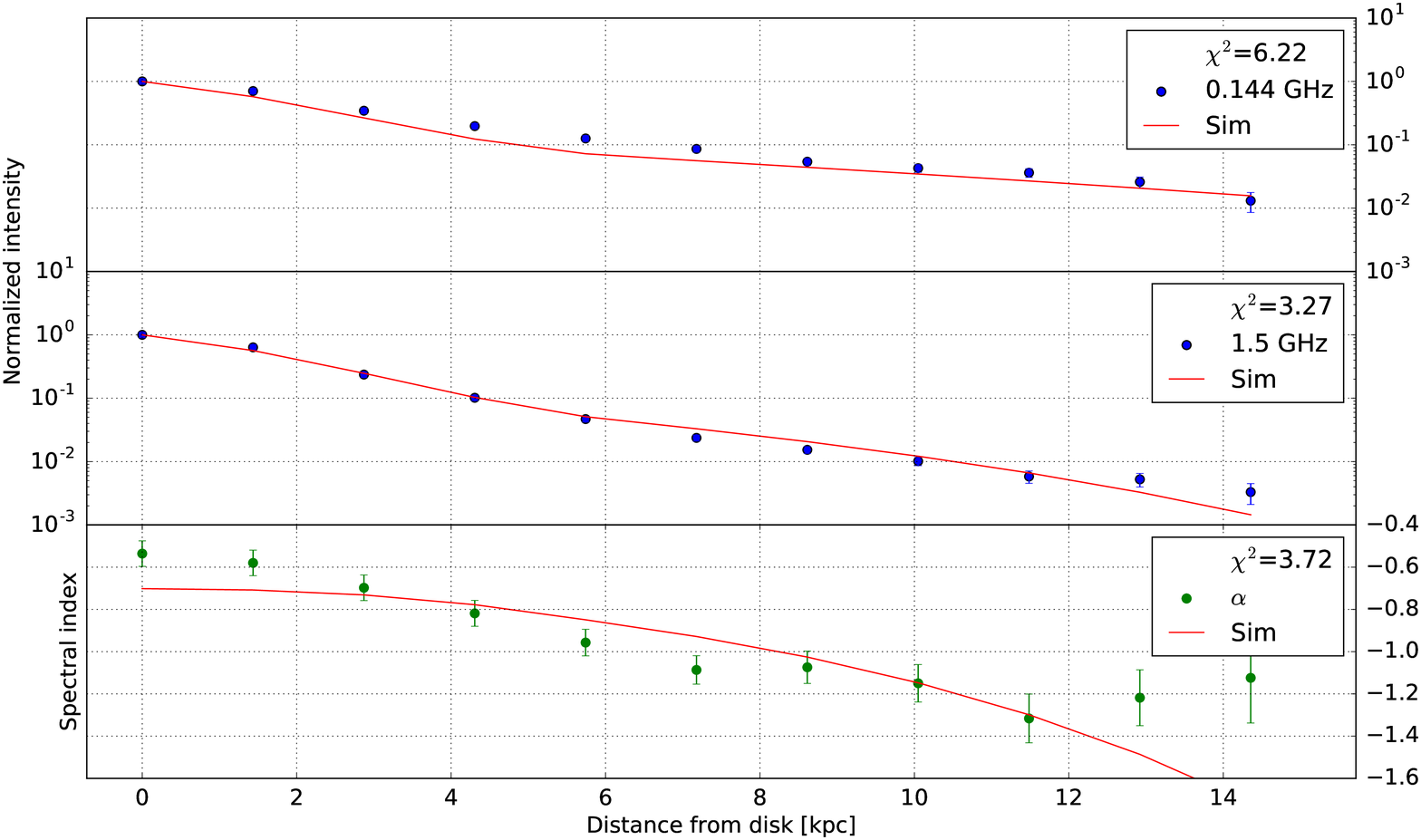}
\caption{Resulting profiles of the diffusive cosmic-ray propagation simulation. Blue dots represent  normalised measured intensities; the top plot shows the LOFAR 144 MHz measurements and middle plot shows the VLA $1.5$ GHz measurements. The bottom panel shows the radio spectral index profile. In all panels, the red lines represent the simulated quantity. The scale for the intensity profiles is logarithmic, whereas the spectral index is shown in linear scale.}
\label{spinplot_diffusion}
\end{figure*}

For a diffusion-type propagation, the model cannot reproduce the observations. With the parameters from Table \ref{Mag_scaleheiths_table}, the following result can be achieved, as presented in Fig.~\ref{spinplot_diffusion}. The resulting reduced $\chi^2$ is $4.4$, which cannot be called an acceptable fit.
The observed spectral index steepens rather gradually from the disc into the halo as a function that can be described as linear. Furthermore, the discovery of a very extended ($\approx$20~kpc) exponential component also hints at advection rather than diffusion, as the latter leads to Gaussian intensity profiles. This is now supported by the findings of our fitting routine. We also tested energy-dependent diffusion coefficients, which we parametrised as $D=D_0(E/1~{\rm GeV})^{\mu}$. We found that this did not improve the fit, but on the contrary made them even worse.

\begin{figure*}[!htbp]
\includegraphics[width=17cm]{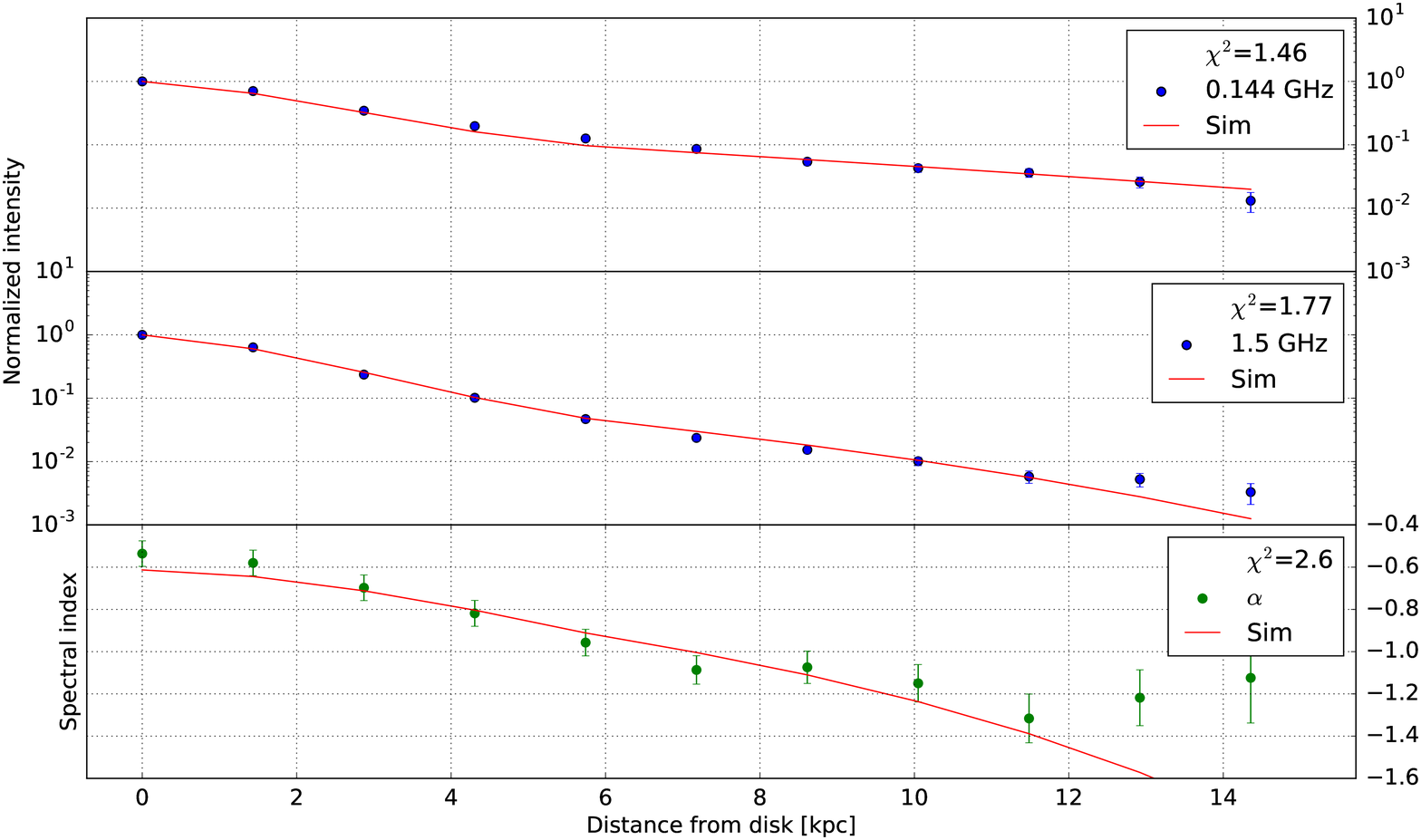}
\caption{Resulting profiles of the advective cosmic-ray propagation simulation with a constant wind speed. Blue dots represent  normalised measured intensities; the top plot shows the LOFAR $144$~MHz measurements and middle plot shows the VLA $1.5$ GHz measurements. The bottom panel shows the radio spectral index profile. In all panels, the red lines represent the simulated quantity. The scale for the intensity profiles is logarithmic, whereas the spectral index is shown in linear scale.}
\label{spinplot_advection}
\end{figure*}

As the next step, we explored an advection model with a constant wind speed. This provides us with a much improved fit with a reduced $\chi^2$ of $1.9$. The resulting best-fitting model is presented in Fig.~\ref{spinplot_advection}, where we find a best-fitting advection speed of $145~\rm km\,s^{-1}$.  
This can be compared with the escape velocity, which, for a truncated isothermal sphere, can be calculated following \cite{Veilleux2005}:
\begin{equation}
v_{\rm{esc}} = \sqrt{2}\, v_{\textnormal{rot}}\, \sqrt[]{1+\ln\left ( \frac{R_{\rm max}}{r}\right )},
\end{equation}
where $\varv_{\textnormal{rot}}$ is the rotational velocity of the galaxy 
(equals to $154~\rm km\, s^{-1}$ for NGC~3556, \citealt{King1997}), 
$r$ is the spherical radial coordinate, and $R_{\rm max}$ is the outer radius of the truncated isothermal sphere. 

We calculated escape velocities assuming three different values for the outer radius  (10, 30, and 60~kpc) and radial distances in the galactic disc ranging from 1 to 10~kpc from the centre of the galaxy. The resulting velocities are presented in Fig.~\ref{windspeed_fig}. The green area corresponds to escape velocities for $R_{\rm max}=10$~kpc; the red area for $R_{\rm max} = 30$~kpc, and the grey area to $R_{\rm max} = 60$~kpc. The lower and upper bounds of each area represent the limits of the radial distance: $1$~kpc at the top and $10$~kpc at the bottom.

\begin{figure*}[!htbp]
\includegraphics[width=17cm]{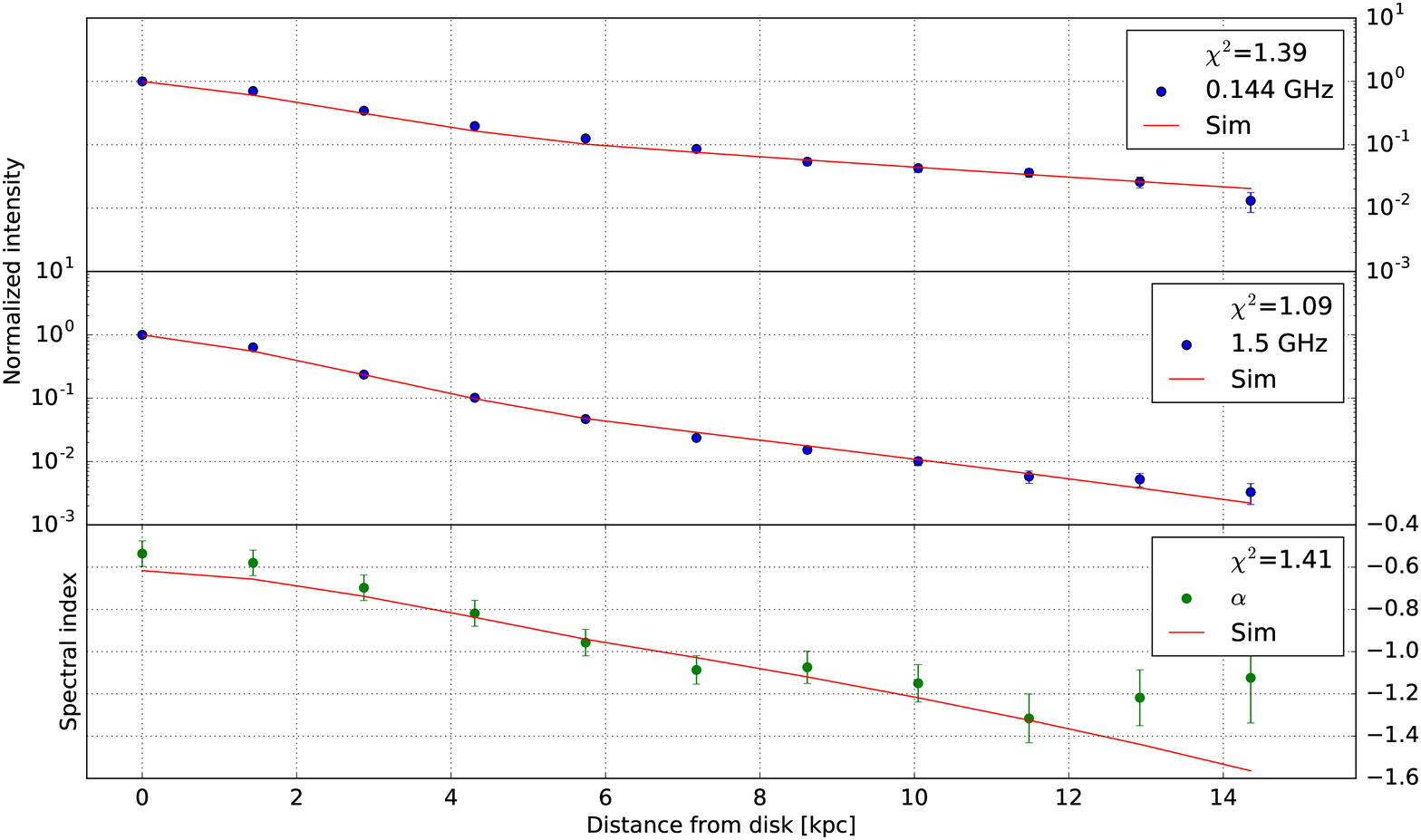}
\caption{Resulting profiles of the advective cosmic-ray propagation simulation using a linearly accelerating wind speed. The blue dots represent  normalised measured intensities; the top plot shows the LOFAR 144 MHz measurements, and middle plot shows the VLA $1.5$ GHz measurements. The bottom panel shows the radio spectral index profile. In all panels, the red lines represent the simulated quantity. The scale for the intensity profiles is logarithmic, whereas the spectral index is shown in linear scale.}
\label{spinplot_best_advection}
\end{figure*}

\begin{figure}[!htbp]
\includegraphics[width=\columnwidth]{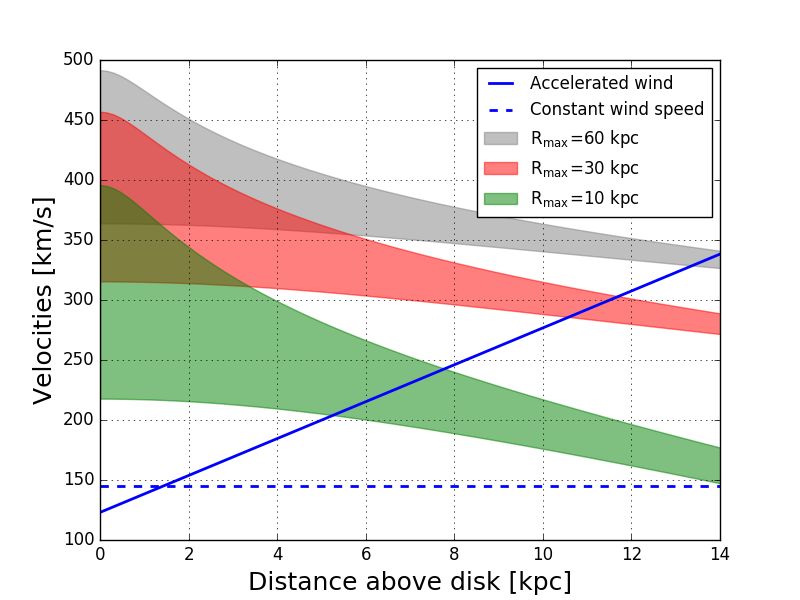}
\caption{Wind speed profile of the different wind models derived from the simulations shown in 
Figs~\ref{spinplot_advection} and
\ref{spinplot_best_advection} (blue lines; constant or linearly increasing speed)
 and ranges of calculated escape velocities for this galaxy. 
 The coloured areas show the escape velocities calculated for different values of the outer radius of the galaxy 
 ($R_{\textrm{max}} = 60$~kpc in grey;  
 $R_{\textrm{max}} = 30$~kpc in red; and 
 $R_{\textrm{max}} = 10$~kpc in green). In each coloured area plot, the radial distance $R$ from the centre of the galaxy in the galactic plane ($R = (r^2-z^2)^{1/2}$, where $r$ is the radius in spherical coordinates and $z$ is the distance above the disc) 
 increases from 1~kpc (top) to 10~kpc (bottom).}
\label{windspeed_fig}
\end{figure}

The fit can be improved if we assume the advection velocity to increase in the halo. We parametrised the advection velocity as
\begin{equation}
\varv(z) = \varv_0 \left (1 + \left (\frac{z}{h_{\varv}}\right )^{\beta}\right ).
\end{equation}
The acceleration parameter, $\beta$, determines the shape of the velocity profile and $h_{\varv}$ is the velocity scale height. We found that $\beta=1$ gave the best-fitting model, which means a linear acceleration. We also tested square root ($\beta = 0.5$) and quadratic ($\beta=2$) velocity profiles. They gave poorer fits. In Fig.~\ref{spinplot_best_advection}, we present the best-fitting accelerated wind model with a reduced $\chi^2$ of $1.25$ with the corresponding advection velocity profile presented in Fig.~\ref{windspeed_fig}. This model starts with a slow advection speed in the midplane of $123~\rm km\,s^{-1}$ but accelerates away from the disc to reach our lowest estimate for the escape velocity at a height of about $5$~kpc. The wind accelerates further so that the advection speed at the outer limit of the observable halo at $z=15$~kpc is $>350~\rm km\,s^{-1}$, reaching the escape velocity of even our largest assumed isothermal sphere 
($R_{max} = 60$~kpc). Figure~\ref{windspeed_fig} shows the simulated wind speed as function of distance from the disc for an accelerated wind model and the escape velocities for different isothermal spheres radii.

For this model the cosmic rays and magnetic field are in approximate energy equipartition even away from the disc in the halo; the cosmic rays slightly dominate the energy density by a factor of up to five. This appears to be a reasonable assumption if  the cosmic rays are able to drive the wind. The best-fitting parameters for our models are presented in Table~\ref{spintable}. 

\begin{table}[!htbp]
\caption{Best-fitting parameters for the cosmic-ray transport models.}

\centering
\begin{tabular}{ccc}
\hline\hline
Parameter & Value &\\
\hline
\multicolumn{3}{c}{Diffusion model}\\
\hline
$B_0$ & $9.0$ & $\mu$G\\
$B_1$ & $5.0$ & $\mu$G\\
$h_{B1}$ & $4.0$ & kpc\\
$h_{B2}$ & $33.0$ &kpc\\
$D_0$ & $21.7$ & $10^{28}~\rm cm^2\,s^{-1}$\\
$\mu$ & $0$\\
$\gamma$ & $2.4$\\
$\chi^2_{\rm red}$ & $4.4$\\
\hline
\multicolumn{3}{c}{Advection model (constant speed)}\\
\hline
$B_0$ & $9.0$ & $\mu$G\\
$B_1$ & $8.0$ & $\mu$G\\
$h_{B1}$ & $3.5$ & kpc\\
$h_{B2}$ & $11.5$ &kpc\\
$\varv_0$ & $145$ & $\rm km\,s^{-1}$\\
$\gamma$ & $2.2$\\
$\chi^2_{\rm red}$ & $1.9$\\
\hline
\multicolumn{3}{c}{Advection model (linearly increasing speed)}\\
\hline
$B_0$ & $9.0$ & $\mu$G\\
$B_1$ & $8.0$ & $\mu$G\\
$h_{B1}$ & $5.5$ & kpc\\
$h_{B2}$ & $21.0$ &kpc\\
$\varv_0$ & $123$ & $\rm km\,s^{-1}$\\
$h_v$ & $8.0$ & kpc\\
$\gamma$ & $2.2$\\
$\chi^2_{\rm red}$ & $1.3$\\
\hline
\label{spintable}
\end{tabular}
\end{table}

\subsection{Small-scale structures in thermal and non-thermal emission} 
\label{sectResultsSmallScaleStruct}
\begin{figure}[!htbp]
{\resizebox{\hsize}{!}{\includegraphics{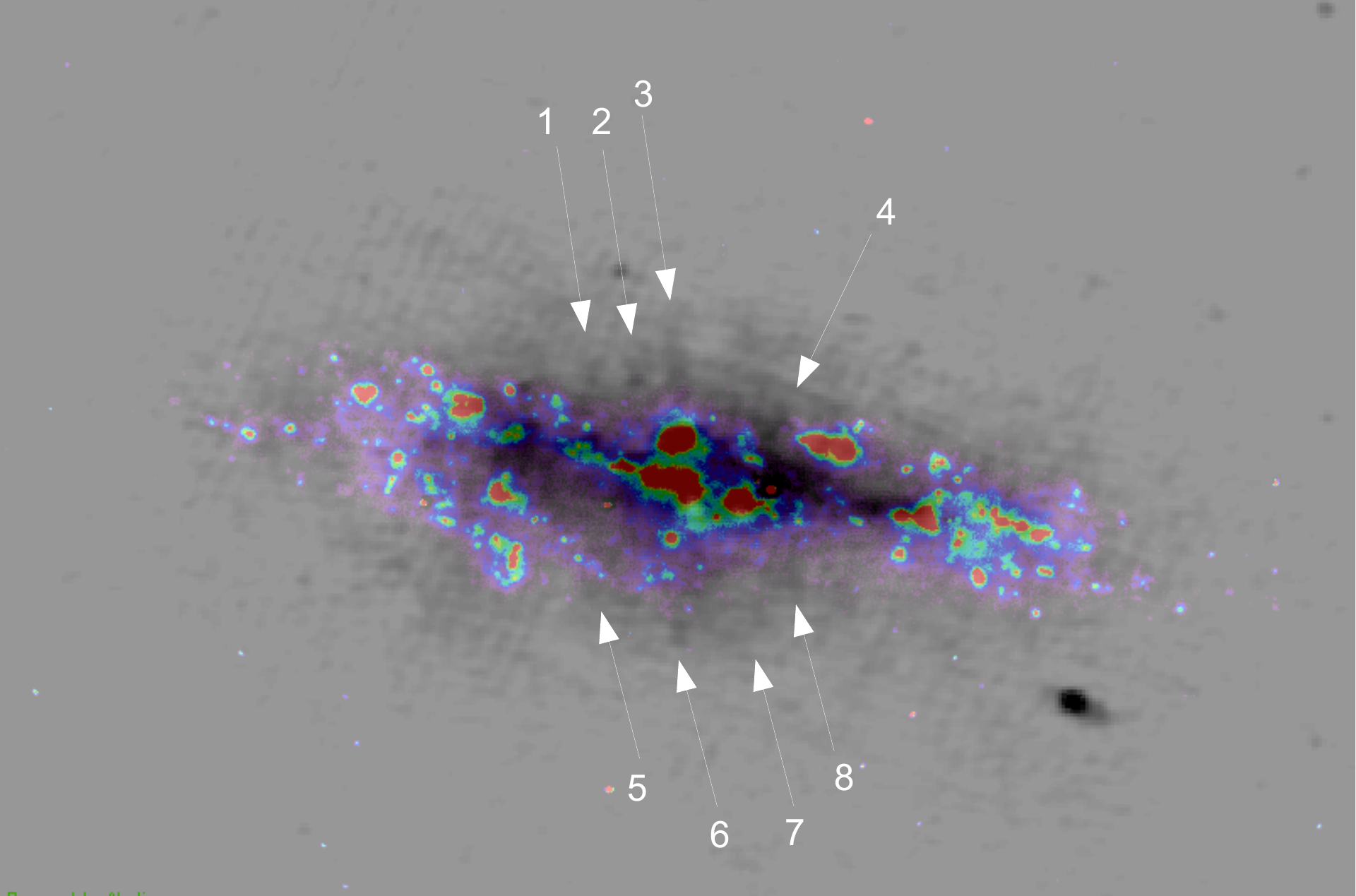}
}}
\caption{
Sub-image measuring $9\arcmin \times 6\arcmin$ of the high-resolution LoTSS
tile containing NGC\,3556 plotted as greyscale image; our continuum-subtracted H$\alpha$ image is overlaid as a pseudo-colour image (intensity scale 
rainbow: from low intensity = violet and blue to high intensity = red). 
North is up, east is left in the image.  Several radio structures are marked 
and numbered.}
\label{HA-LoTSS}
\end{figure}

Since the problems of the \textsc{KillMS} pipeline, negative bowls and possibly missing flux were not noticeable at high 
surface brightness features, we used the high-resolution $5\arcsec$ 
LoTSS image of NGC\,3556 to compare the structure of the non-thermal 
radio continuum with the thermal emission traced by the continuum-subtracted 
H$\alpha$ image. 
For this, we overlaid the LoTSS high-resolution images as greyscales with 
a pseudo-colour version of the continuum-subtracted H$\alpha$ image. The result 
is given in Fig.~\ref{HA-LoTSS}.  
Several aspects are worth noting. Our H$\alpha$ image shows some extended diffuse
H$\alpha$ emission
in addition to the \ion{H}{ii} regions, as already noted by
\citet{Collins2000}. 
While there is a general correlation 
between H$\alpha$ and radio continuum, there is not a good one-to-one correspondence 
of the brightest \ion{H}{II} regions and the brightest LOFAR emission neither
spatially nor in flux/intensity. This may be an effect of the dust
absorption, which affects the H$\alpha$ line flux in the disc. The number 
of \ion{H}{II} regions that coincide with LOFAR emission are therefore at the 
front side of the galaxy. The H$\alpha$ bright central region is also the brightest 
region in radio continuum, implying efficient dust removal, probably due to 
a wind from this central region.   
There are several kiloparsec-sized radial filaments 
visible in the LoTSS image extending into the halo; see arrow marks in 
Fig.~\ref{HA-LoTSS}. In most cases (\#2, 3, 4, and 7) these filaments connect back to 
large \ion{H}{ii} regions in the disc of NGC\,3556, or broader filaments (\#1, 5, 6, and 8) 
connect to a large, bright region containing DIG.
It is tempting to identify these filaments as magnetised chimneys from giant \ion{H}{ii} regions 
and more evolved, bright regions of DIG with still significant SNe activity. Despite the lower 
resolution and smaller signal-to-noise ratio, the overall morphology of the radial filaments extending into the halo resembles that of the 
starburst galaxy M82, see for example\ the unpublished 5 GHz map (NRAO News Release, 2014 February 3).  
The non-detection of a diffuse large-scale outflow in our H$\alpha$ data is most probably due to the relatively low sensitivity of our optical 
data.  
Still, the clear connection of non-thermal radio continuum filaments to large star-forming 
regions and regions of large diffuse H$\alpha$ emission implies an outflow/wind 
driven by the combined pressure from cosmic rays and hot thermal gas as result of SNe activity.

\section{Discussion}
\label{sec:discussion}
As shown in Sect.~\ref{sec:results}, the extent of the faint halo increases from high to low frequency. The radial extent, however, does not increase. This has been observed in many spiral edge-on galaxies (e.g.\ in the CHANG-ES sample) and can be interpreted as a cosmic-ray driven galactic wind \citep{Butsky2018}. 

Furthermore, the extent in the CHANG-ES maps is generally smaller than in the LOFAR map. This can be a result of the missing-spacing problem, where an interferometer does not `see' emission on angular scales above a certain size. For the D-array/$C$-band combination, however, the largest angular scale is about $4\arcmin$ in size.\footnote{\url{https://science.nrao.edu/facilities/vla/\\docs/manuals/oss/performance/resolution}} 
This is comparable to the extent seen in the LOFAR map, so it is reasonable to assume that the halo would be seen in its entirety. Hence, we conclude that since the observed halo extent is smaller than in the $144$~MHz map, the CRe did not propagate any further as a result of spectral ageing.

As the top panel of Fig.~\ref{figPIBRM} 
shows, the galaxy exhibits polarised $1.5$~GHz emission in only one patch localised at the eastern side of the galaxy. Such localised regions of PI have been observed before, such as in NGC\,5055 \citep{Heald2009}. This side of the galaxy is the approaching side \citep{Wiegert2011}. \cite{Braun2010} argued that such an azimuthal asymmetry of PI can be the result of a combination of Faraday depolarisation and a field projection in which a spiral disc field is combined with a quadrupolar halo field. Since this depends on the presence of Faraday depolarisation, the PI should be found across the galaxy's disc at higher frequency observations. We can test with our 6 GHz observations.

The middle panel of Fig.~\ref{figPIBRM} shows that polarisation can indeed be found across the disc at $6$~GHz, suggesting that the magnetic topology proposed by \cite{Braun2010} may be present in this galaxy. An argument for this can also be found in the polarisation vectors found in both maps. While the $L$-band map only shows a small patch with a curvature that resembles a quadrupolar field, the higher resolution $C$-band map shows more locations, which resemble the outline of a quadrupolar field. First, there are vertical field lines in the centre of the galaxy, and second, on both sides of the galaxy there are plane parallel field lines that curve upwards and downwards, as one expects from a quadrupolar field. The reason why the parallel part of the field is missing in the $L$-band map is that the polarised emission comes from a foreground layer of the galaxy, while emission from a higher depth along the line of sight is depolarised. 

Figure~\ref{figPIBRM} (bottom panel)
indicates that the magnetic field on the southern side of the galaxy is pointing away from the observer and the northern side of the magnetic field of the galaxy is pointing towards the observer. This would contradict a quadrupole field, since the magnetic field in such a field configuration would have the same direction above and below the centre of the field. Unfortunately, the Faraday resolution in $C$ band is very poor, 
$\approx 1000~\rm rad\,m^{-2}$ because of the observing set-up of the $C$-band observations used by CHANG-ES, such that the values obtained from the $C$-band map do not provide  us with any further useful information.

The fitted scale heights at two frequencies can be used to approximate the CRe propagation type, by calculating the ratio of the scale heights because they are approximately proportional to the ratio of the frequencies. The relevant relations for different propagation types and their derivations are found, for example, in \cite{changes9} as follows:
\begin{eqnarray}
\textnormal{Diffusion:~} \dfrac{h_1}{h_2} = \left(\dfrac{\nu_1}{\nu_2}\right)^{-1/8}\\
\textnormal{Energy-dependent diffusion ($\mu=0.5$):~} \dfrac{h_1}{h_2} = \left(\dfrac{\nu_1}{\nu_2}\right)^{-1/4}\\
\textnormal{Advection:~}: \dfrac{h_1}{h_2} = \left(\dfrac{\nu_1}{\nu_2}\right)^{-1/2} \, .
\end{eqnarray}

These loss processes are all for a synchrotron loss dominated halo, which means that the CRe lose their energy through synchrotron radiation before they can escape the galaxy. The three ratios are for energy-independent diffusion, energy-dependent diffusion, and advection. However, for an escape-dominated halo, in which the CRe propagate fast enough so that they can escape the galaxy before they radiate away their energy, the ratio can be smaller or even approaching unity.
For this work, the expected ratios for $\nu_1/\nu_2 = $144$~{\rm MHz} / $1.5$~{\rm GHz}$ are given in Table~\ref{ratiotable}.

\begin{table}[!htbp]
\caption{Expected scale height ratios given the frequencies used in this work for different propagation types, based on \cite{changes9}.}
\begin{center}
\begin{tabular}{cccc}
\hline\hline
Diffusion & Diffusion ($\mu=0.5$) & Advection\\
\hline
$1.34$ & $1.80$ & $3.24$ \\
\hline
\label{ratiotable}
\end{tabular}
\end{center}
\end{table}

The measured ratio of the scale heights at 144~MHz to at $1.5$~GHz is $1.3 \pm 0.41$ for the thin disc and $1.81\pm 0.71$ for the thick disc. The large errors are the result of the scale height errors that propagate through the calculation. Therefore, this calculation is only conclusive in that sense that an advection-type propagation in a synchrotron loss dominated halo can be excluded.

The CRe propagation simulations, however, are more conclusive and show that only advection results in a good fit, such as the model with a constant wind speed of $145~\rm km\,s^{-1}$. An even better fit was achieved using a linearly accelerated wind. This then yields an initial wind speed of $\varv_0 = 123~\rm km\,s^{-1}$. In this model, the advective timescale with the wind profile shown in Fig.~\ref{windspeed_fig} is $67~\rm Myr$ at the edge of the observable halo, which is the time that the CRe would need to reach a height of $15$~kpc. Depending on the assumed outer radius of the isothermal sphere, the escape velocity is reached at heights between $5$ and $15$~kpc. This is similar to values obtained by \mbox{\cite{Everett2008}} and \cite{Mao2018} who simulated cosmic-ray driven outflows.
To calculate if it is possible for the CRe to escape the galaxy, we calculated the energy of CRe observed at LOFAR frequencies and their synchrotron lifetime.

The energy of CRe can be approximated by the relation
\begin{equation}
E = \left(\frac{\nu}{16.1\,{\rm MHz}} \right)^{0.5} 
\left(\frac{B_\perp}{\mu G}\right)^{-0.5},
\end{equation}
since the radiation of CRe is mostly synchrotron radiation, which depends on the initial energy and the strength of the magnetic field \citep{changes9}. Calculating this energy using the LOFAR observing frequency of 144~MHz and a magnetic field strength of $9~\mu$G, the lower limit of the values found in the centre of the galaxy, the resulting energy is 1~GeV.

Using equations from \cite{heesen2009} and \cite{changes9}, the synchrotron lifetime can now be calculated using

\begin{equation}
\frac{t_{\rm syn}}{{\rm yr}} = 8.35 \times 10^{9} 
\left( \frac{E}{{\rm GeV}} \right)^{-1} 
\left( \frac{B_\perp}{\mu G} \right)^{-2}  \, . 
\end{equation}

With a magnetic field strength of $9~\mu$G, the field strength we assumed for the cosmic-ray propagation simulations in the disc plane, and a CRe energy of 1~GeV, the resulting synchrotron lifetime is of the order of 100~Myr. The calculated value for the synchrotron lifetime is higher than the value obtained from the accelerated wind model, therefore we can conclude that the CRe can indeed escape from the galaxy before they are lost to observations owing to their synchrotron lifetime.

\begin{figure}[!htbp]
\includegraphics[width=\columnwidth]{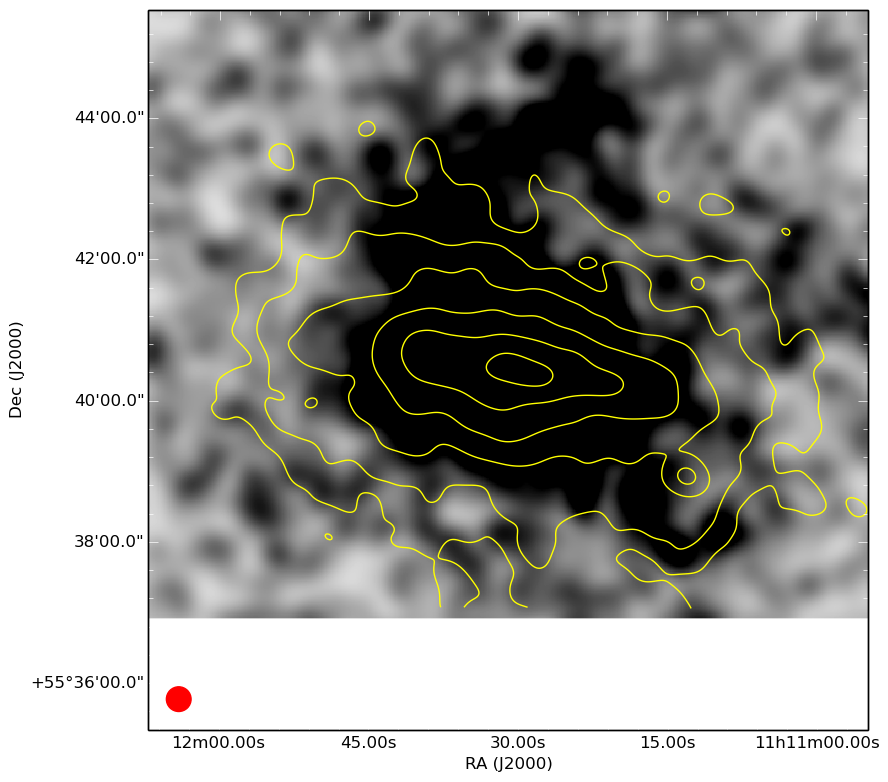}
\caption{LOFAR 144 MHz radio continuum intensity contours overlaid onto a \emph{Suzaku} soft X-ray ($0.3$--2~keV) image. The contours start at $600~\mu\rm Jy\,beam^{-1}$ and increase by factors of two. The filled red circle represents the 21\arcsec FWHM synthesised beam.}
\label{fig:suzaku}
\end{figure}

Another sign of a wind can be found in the \emph{Suzaku} X-ray observations of this galaxy, as shown in Fig.~\ref{fig:suzaku}. 
This image is constructed in the $0.3$--2~keV band with a $67.5$ ks observation of the X-ray Imaging Spectrometer, XIS-3, aboard the observatory. The map was smoothed to a resolution of $21\arcsec$ FWHM.

The lack of detailed correlation between the X-ray and radio emission, however, indicates the outflows driven by the hot gas and cosmic rays may have very different dynamics, which are yet to be investigated. These results led us to the conclusion that the measured values of the magnetic field strength and the fitted propagation parameters fit well into simulations for this type of a galaxy, which indeed show the existence of a galactic wind for such a galaxy.

\section{Summary and conclusions}
\label{sec:conclusions}
In this paper, we have utilised deep LOFAR observations from the LoTSS survey at $144~\rm MHz$ to study the nearby late-type spiral galaxy NGC~3556. Its high inclination angle means that we can study the galaxy in an edge-on position, giving us a good view of its halo. We used further VLA observations from the CHANG-ES survey at effective frequencies of $1.5$ and 6~GHz to calculate radio spectral indices. We performed RM synthesis to measure the linearly polarised emission to study the structure of the magnetic field. Furthermore, we used 1D cosmic-ray transport models applied to the electrons, assuming a steady state with a balance between injection and energy losses. These are our main results:

\begin{itemize}
\item We obtained polarisation and rotation measure maps through RM synthesis, which show that the PI is only present on the eastern, approaching side of the galaxy.
Polarisation maps in $C$ band, however, show the outline of what could be a quadrupolar magnetic field. This would explain the missing polarisation in $L$ band via a magnetic topology that has been proposed by \cite{Braun2010}.
\item We analysed the total power maps using stripe integration techniques and found a scale height of $1.43$~kpc for the thin disc at $1.5$~GHz and $1.86$~kpc at 144~MHz. For the thick disc, the scale heights are $3.28$ and $5.93$~kpc, respectively.
\item Using the equipartition assumption and an exponential fit to the magnetic field profile, we calculated the magnetic field strength in the galaxy to be $9~\mu$G with a magnetic field scale height $1.5$~kpc for the thin disc, and $23.7$~kpc for the thick disc. 
\item We simulated cosmic-ray propagation using 1D transport models for pure diffusion and advection. We found that we can rule out diffusion as the dominating transport process since advection gives much better fits to the data. Our best-fitting model is a linearly accelerating wind with an initial wind speed of $123~\rm km\,s^{-1}$ that reaches the escape velocity at a height between 5 and $8$~kpc if a truncation radius of $10$~kpc is assumed, between $10$ and $12$~kpc if a truncation radius of $30$~kpc is assumed, and finally, at $\approx$ $15$~kpc if a radius of $60$~kpc is assumed. In such a model, the cosmic rays and magnetic field are close to energy equipartition in the halo with the cosmic rays dominating by a factor of a few.
\item We estimate the CRe lifetime at 1~GeV as seen by LOFAR to be approximately 100~Myr before they radiate their energy away. From the cosmic-ray propagation model we calculated that the CRe advective timescale is 67~Myr to reach a height of 15~kpc, the sensitivity limit of our observations. Hence, this model is consistent with the CRe lifetime estimate.

\end{itemize}

Our work demonstrates the potential that LOFAR has in studying the radio haloes of nearby galaxies. Our discovery of a very extended halo magnetic field component with a scale height of at least 20~kpc shows that we have access to components in galaxies that thus far have evaded detection at GHz frequencies. Our best-fitting model requires an accelerating wind, which, to our knowledge, is also a first for modelling of radio haloes in external galaxies. Our advection solution starts with a modest advection speed of $123~\rm km\,s^{-1}$ to accelerate and reach the escape velocity a few kiloparsec away from the disc. Such a behaviour is predicted by the many models of cosmic-ray driven winds that have recently gained traction. Radio continuum observations offer opportunities to measure the transport of cosmic rays outside of the Milky Way to test these models. Since LoTSS will observe the entire northern hemisphere, we can be sure that many more exciting opportunities will arise to study this subject further. More specifically, the variation of cosmic-ray driven wind properties with underlying SFRD or other galaxy properties can be explored, which is unique and powerful to constrain models.

\begin{acknowledgement}
The data used in work was in part processed on the Dutch national e-infrastructure with the support of SURF Cooperative through grant e-infra 160022 \& 160152.\\
\newline
BNW acknowledges support from the Polish National Centre of Sciences (NCN), grant no. UMO-2016/23/D/ST9/00386.\\
\newline
This paper is based (in part) on data obtained with the International LOFAR
Telescope (ILT) under project code LC3\_008. LOFAR (van Haarlem et al. 2013) is the LOw
Frequency ARray designed and constructed by ASTRON. It has observing, data
processing, and data storage facilities in several countries, which are owned by
various parties (each with their own funding sources) and are collectively
operated by the ILT foundation under a joint scientific policy. The ILT resources
have benefitted from the following recent major funding sources: CNRS-INSU,
Observatoire de Paris, and Université d'Orléans, France; BMBF, MIWF-NRW, MPG,
Germany; Science Foundation Ireland (SFI), Department of Business, Enterprise and
Innovation (DBEI), Ireland; NWO, The Netherlands; The Science and Technology
Facilities Council, UK; Ministry of Science and Higher Education, Poland.\\
\newline
Part of this work was carried out on the Dutch national e-infrastructure with the support of the SURF Cooperative through grant e-infra 160022 \& 160152.  The LOFAR software and dedicated reduction packages on \url{https://github.com/apmechev/GRID\_LRT} were deployed on the e-infrastructure by the LOFAR e-infragroup, consisting of J. B. R. Oonk (ASTRON \& Leiden Observatory), A. P. Mechev (Leiden Observatory) and T. Shimwell (ASTRON) with support from N. Danezi (SURFsara) and C. Schrijvers (SURFsara).\\
\newline
This research has made use of data analysed using the University of
Hertfordshire high-performance computing facility
(\url{http://uhhpc.herts.ac.uk/}) and the LOFAR-UK computing facility
located at the University of Hertfordshire and supported by STFC
[ST/P000096/1].\\
\newline
Funding for the Sloan Digital Sky Survey IV has been provided by the Alfred P. Sloan Foundation, the U.S. Department of Energy Office of Science, and the Participating Institutions. SDSS acknowledges support and resources from the Center for High-Performance Computing at the University of Utah. The SDSS website is \url{www.sdss.org}.

SDSS is managed by the Astrophysical Research Consortium for the Participating Institutions of the SDSS Collaboration including the Brazilian Participation Group, the Carnegie Institution for Science, Carnegie Mellon University, the Chilean Participation Group, the French Participation Group, Harvard-Smithsonian Center for Astrophysics, Instituto de Astrofísica de Canarias, The Johns Hopkins University, Kavli Institute for the Physics and Mathematics of the Universe (IPMU) / University of Tokyo, the Korean Participation Group, Lawrence Berkeley National Laboratory, Leibniz Institut für Astrophysik Potsdam (AIP), Max-Planck-Institut für Astronomie (MPIA Heidelberg), Max-Planck-Institut für Astrophysik (MPA Garching), Max-Planck-Institut für Extraterrestrische Physik (MPE), National Astronomical Observatories of China, New Mexico State University, New York University, University of Notre Dame, Observatório Nacional / MCTI, The Ohio State University, Pennsylvania State University, Shanghai Astronomical Observatory, United Kingdom Participation Group, Universidad Nacional Autónoma de México, University of Arizona, University of Colorado Boulder, University of Oxford, University of Portsmouth, University of Utah, University of Virginia, University of Washington, University of Wisconsin, Vanderbilt University, and Yale University.\\
\newline
The work at Ruhr-University Bochum is supported by BMBF Verbundforschung
under D-LOFAR IV - FKZ: 05A17PC1.\\
\newline
The project has in part also benefitted from the exchange programme between  Jagieollonian University Krakow and Ruhr-University Bochum.\\

\newline
We thank Olaf Wucknitz for his useful comments.

\newline
We thank the anonymous referee for  constructive and helpful comments.
\end{acknowledgement}

\bibliographystyle{aa}
\bibliography{bib}
\end{document}